%% file: elsarticle-template-num.tex
\documentclass[preprint,12pt]{elsarticle}

\usepackage{amssymb}
\usepackage{hyperref}
\usepackage{graphicx}
\usepackage{subfigure}
\usepackage{epstopdf}
\usepackage{color}
\usepackage{amsmath}
\usepackage{amsthm}
\usepackage[noend]{algpseudocode}
\usepackage{booktabs}
\usepackage{algorithmicx,algorithm}
\usepackage{multirow}

\journal{}

\begin{document}

\begin{frontmatter}


\title{Improving Low-Resource Knowledge Tracing Tasks by Supervised Pre-training and Importance Mechanism Fine-tuning}
\affiliation[a]{organization={Shenzhen International Graduate School, Tsinghua University},
            city={Shenzhen},
            postcode={518055},
            country={China}}
\affiliation[b]{organization={Guangdong Institute of Smart Education, Jinan University},
            city={Guangzhou},
            postcode={510610},
            country={China}}
\affiliation[c]{organization={TAL Education Group},
            city={Beijing},
            postcode={100080},
            country={China}}
\author[a]{Hengyuan Zhang}
	\ead{hengyuan.zhang88@gmail.com}
	\author[b]{Zitao Liu\corref{cor1}}
	\ead{zitao.jerry.liu@gmail.com}
	\author[c]{\textcolor[RGB]{0,0,1}{Shuyan Huang}}
    \author[a]{\textcolor[RGB]{0,0,1}{Chenming Shang}}
	\author[b]{\textcolor[RGB]{0,0,1}{Bojun Zhan}}
	\author[a]{\textcolor[RGB]{0,0,1}{Yong Jiang}}

    \cortext[cor1]{Corresponding author}

\begin{abstract}
    Knowledge tracing (KT) aims to estimate student's knowledge mastery based on their historical interactions. Recently, the deep learning based KT (DLKT) approaches have achieved impressive performance in the KT task. These DLKT models heavily rely on the large number of available student interactions. However, due to various reasons such as budget constraints and privacy concerns, observed interactions are very limited in many real-world scenarios, a.k.a, low-resource KT datasets. Directly training a DLKT model on a low-resource KT dataset may lead to overfitting and it is difficult to choose the appropriate deep neural architecture. Therefore, in this paper, we propose a low-resource KT framework called \emph{LoReKT} to address above challenges. Inspired by the prevalent ``pre-training and fine-tuning'' paradigm, we aim to learn transferable parameters and representations from rich-resource KT datasets during the pre-training stage and subsequently facilitate effective adaptation to low-resource KT datasets. Specifically, we simplify existing sophisticated DLKT model architectures with purely a stack of transformer decoders. 
    We design an encoding mechanism to incorporate student interactions from multiple KT data sources and develop an importance mechanism to prioritize updating parameters with high importance while constraining less important ones during the fine-tuning stage. We evaluate \emph{LoReKT} on six public KT datasets and experimental results demonstrate the superiority of our approach in terms of AUC and Accuracy. To encourage reproducible research, we make our data and code publicly available at \url{https://github.com/rattlesnakey/LoReKT}.
\end{abstract}



\begin{keyword}
    Educational data mining \sep
    Knowledge tracing \sep
    Pre-training and fine-tuning \sep
    Importance mechanism



\end{keyword}

\end{frontmatter}

\section{Introduction}
\label{sec:intro}
\input{intro.tex}

\section{Related Work}
\label{sec:related}
\input{related.tex}

\section{Problem Statement}
\label{sec:problem_statement}
\input{problem_statement.tex}

\section{The Framework}
\label{sec:method}
\input{method.tex}

\nocite{kong2022multitasking,li2023multi}

\section{Experiment}
\label{sec:experiment}
\input{exp.tex}

\section{Conclusion}
\label{sec:conclusion}
\input{conclusion.tex}

\nocite{shang2024understanding,zhang2022fine,kong2022blcu}

\section*{CRediT authorship contribution statement}
\par{Hengyuan Zhang: Methodology, Conceptualization, Investigation, Writing.}
\par{Zitao Liu: Writing - review \& editing, Supervision.}
\par{Shuyan Huang: Writing - review \& editing, Supervision.}
\par{Chenming Shang: Investigation, Methodology, Writing.}
\par{Bojun Zhan: Investigation, Writing.}
\par{Yong Jiang: Supervision.}

\section*{Declaration of competing interest}
\par{The authors declare that they have no known competing financial interests or personal relationships that could have appeared to influence the work reported in this paper.}

\bibliographystyle{elsarticle-num-names}
\bibliography{myKBS}
\end{document}

%% file: intro.tex
\begin{figure}[!thbp]
    \centering
    \includegraphics[width=0.8\columnwidth]{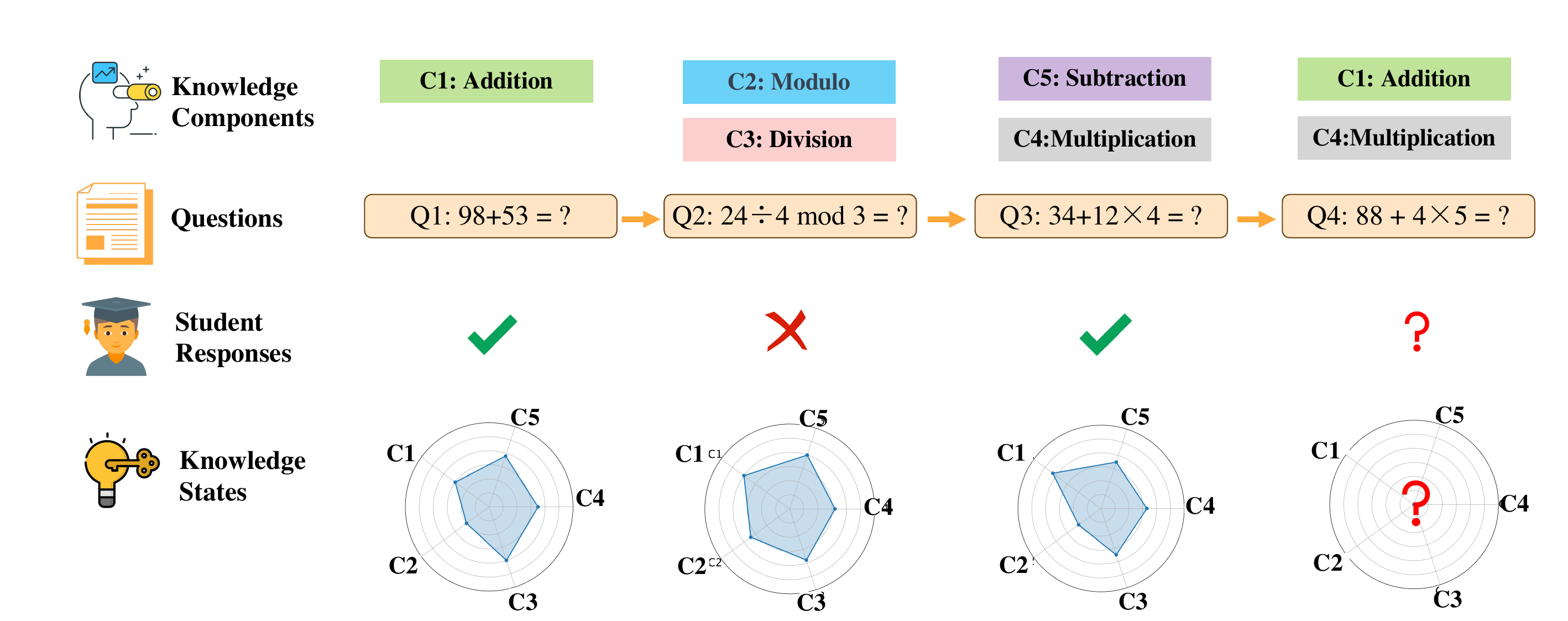}
    \caption{An illustration of the KT problem.}
    \label{fig:kt_illustration}
\end{figure}

Knowledge tracing holds a pivotal role within the realm of Intelligent Tutoring Systems (ITS)~\citep{zhang2018three,su2021time,song2022bi,ke2024hitskt}. 
Its primary objective is to forecast students' performance on questions by estimating their mastery of individual knowledge components (KCs\footnote{A KC is a generality of everyday terms like concept, principle, or skill.}) through an analysis of their past interactions.
A KC is a description of a mental structure or process that a learner uses, alone or in combination with other KCs, to accomplish steps in a task or a problem. 
Take Figure~\ref{fig:kt_illustration} as an example.
The student has successively responded to four questions ($Q_1$ to $Q_4$), achieving correct answers for $Q_1$ and $Q_3$, while $Q_2$ is answered incorrectly. 
This pattern suggests that the student may have a proficient understanding of the ``Addition'', ``Subtraction'', and ``Multiplication'' KCs, but lacks familiarity with the ``Modulo'' and ``Division'' KCs. 
Leveraging the current knowledge mastery, the KT task aims to predict the student's performance on the upcoming sixth question, $Q_4$.
After gaining insights into students' knowledge mastery through KT, educators can promptly pinpoint weaknesses and provide targeted exercises for improvement.
Additionally, this information can assist online learning platforms in providing a series of adaptive learning services such as learning resource recommendations, customizing student learning paths, and personalizing teaching plans~\citep{liu2019ekt,wu2020exercise}.

Recently, with the remarkable progress of deep learning techniques, many studies develop deep learning based KT (DLKT) models that are trained on massive students' historical interactions to pursue high accuracy on students' knowledge mastery estimations. 
Thus, many publicly available educational datasets have been released for training an effective DLKT model. 

However, due to the users' privacy protection of the educational applications and the different learning energy and enthusiasm of students, it is extremely difficult to collect large-scale high-quality student interaction sequences from real-world educational environments. 
Therefore, the educational datasets for KT model training frequently involve limited student learning records. 
For example, ASSISTments2009 is one of the classical KT datasets and the observed interaction records are only collected from 4,217 students. 
However, most of the state-of-the-art DLKT models are designed with stacks of neural networks such as recurrent neural networks, memory networks \citep{zhang2017dynamic,nagatani2019augmenting,sonkar2020qdkt,guo2021enhancing}. 
Directly training a DLKT model on such a low-resource KT dataset is very easy to run into the problem of overfitting. 
Furthermore, it is unclear what type of model architectures are most suitable for low-resource KT datasets in previous KT works. 

To enhance the learning capability from low-resource datasets of deep learning-based models, some studies perform ``pre-training and fine-tuning'' paradigm \citep{zoph2016transfer,tu2019end,brown2020language,gira2022debiasing}.
This paradigm leverages rich-resource datasets to pre-train a model first and transfer the learned parameters to the low-resource dataset. 
Motivated by these promising studies, we propose a simple yet effective framework called \emph{LoReKT}.
\emph{LoReKT} aims to improve the performance on the low-resource KT dataset by transferring the knowledge tracing capability from the model pre-trained on multiple rich-resource KT datasets.
More specifically, in the pre-training stage, we build a foundational pre-trained KT model using a stack of transformer decoders based on multiple rich-resource KT datasets. 
To enhance the model's capacity for integrating information from both questions and concepts, we introduce data type embeddings.
Furthermore, to enable the model to learn the distinct and shared tracing patterns from multiple KT datasets, we introduce dedicated dataset embeddings for each KT dataset.
In the fine-tuning stage, we propose an importance vector-based fine-tuning strategy to allow the model to focus on updating crucial parameters for the specific target low-resource dataset while constraining unimportant parameters to prevent the learning and memorization of noisy information.

As a result, the implementation of \emph{LoReKT} has the following merits:
\begin{itemize}
    \item The framework avoids direct training by initially pre-training the model on rich-resource KT datasets and subsequently fine-tuning it on specific low-resource KT datasets. This approach mitigates the risk of overfitting.
    \item The framework leverages a stack of transformer decoders as its backbone, which has demonstrated excellent performance in various ``pre-training and fine-tuning'' scenarios. Moreover, this backbone requires no additional architecture design efforts, simplifying the process and reducing the reliance on specific model architectures for particular datasets.
    \item To further mitigate the overfitting issue, the framework employs a fine-tuning strategy based on an importance mechanism, restricting the learning of less important parameters to prevent the memorization of noisy information.
    \item To ensure that our approach can be fairly comparable with other recently developed DLKT models, we follow a publicly available standardized KT task evaluation protocol \citep{liu2022pykt}. 
    We conduct comprehensive and rigorous experiments on three public rich-resource datasets and three low-resource datasets.
    The results show that after pre-training, the pre-trained KT model comes close to the performance of previous approaches on rich-resource datasets. 
    By fine-tuning the pre-trained KT model on the low-resource datasets, it achieves superior prediction performance in terms of AUC and Accuracy compared to 17 baselines.    
\end{itemize}

%% file: related.tex
\subsection{Deep Learning based Knowledge Tracing}
Deep Knowledge Tracing (DKT) has pioneered the application of deep learning in knowledge tracing tasks by employing a Long Short Term Memory (LSTM) layer to encapsulate students' knowledge states and predict students' response performances \citep{piech2015deep}. 
Since then, many methods tend to use deep learning techniques to solve KT problem \citep{zhang2018three,su2021time,song2022bi,ke2024hitskt,zhang2017dynamic,yeung2018addressing,nagatani2019augmenting,lee2019knowledge,pandey2019self,ghosh2020context,tiana2021bekt,song2022survey,guo2021enhancing,ma2022spakt,liu2023simplekt,zhang2024question}.
For example,
\citet{yeung2018addressing} leveraged prediction-consistent regularization mechanism to mitigate issues related to input reconstruction failure and prediction inconsistency in the context of DKT~\cite{piech2015deep}.
\citet{zhang2017dynamic} integrated a meticulously designed static key matrix for storing the interconnections among different knowledge components (KCs). Simultaneously, it utilizes a dynamic value matrix to iteratively update the knowledge state of students.
Motivated by the learning curve theory~\cite{learning_curve},~\citet{nagatani2019augmenting} took student's forgetting behavior into consideration to enhance DKT~\cite{piech2015deep}. 
\citet{lee2019knowledge} used student knowledge state encoder and skill encoder to predict the student response performance via the dot product.
\citet{tiana2021bekt} performed multi-task learning based on the bidirectional encoder representations to construct mixed representations of questions.
To mitigate the potential issue of limited generalization in DLKT, adversarial training techniques, such as adversarial perturbations, are introduced to the original student interaction sequence.
Specifically, \citet{guo2021enhancing} improveed the generalization capability of the DLKT model by incorporating adversarial perturbations at the embedding level of the student interaction sequence. 
The carefully designed perturbations contribute to the model's effective generalization across diverse student interactions.
Moreover, certain studies have concentrated on exploring the interactions between student responses and questions, as well as the associations between questions and KCs.
For example, \citet{nakagawa2019graph} constructed a question-concept knowledge graph and utilized graph neural network to aggregate the node features related to the corresponding concepts and subsequently updates the student's knowledge states effectively.
Additionally, \citet{ma2022spakt} employed self-supervised learning paradigm to identify the latent relationship between questions and KCs, thereby enhancing input representations.
Another research direction focused on the interdependence among student interactions, aiming to capture finer details embedded within them.
For instance, \citet{pandey2019self} utilized a self-attention mechanism to grasp the relationships between exercises and students' responses.  
\citet{ghosh2020context} presented AKT, which utilizes two self-attention modules to extract the inner relevance of questions and interactions respectively, and explicitly model students' forgetting behaviors via a monotonic attention mechanism. 

In this paper, unlike the aforementioned DLKT methods that are committed to developing a series of sophisticated architectures, our \emph{LoReKT} is based on a stack of simple transformer decoders. 
This unified backbone aims to break down disciplinary barriers and learn consistent representations across multiple KT datasets.

\subsection{Pre-training for Low-resource Setting}
In real-world scenarios, encounters with low-resource settings are commonplace, and the paradigm of "pre-training and fine-tuning" has consistently proven its efficacy in addressing challenges within such contexts \citep{yosinski2014transferable,shang2019pre,bansal2019pre,zhang2021pdaln,liu2022improved,10103146,zhang-etal-2023-assisting}.
For example, 
\citet{yosinski2014transferable} explored the transferability of AlexNet and observed that the initial three layers of AlexNet encapsulate general features conducive to transferability.
By introducing fine-tuning to the neural network, it successfully mitigated data variability and scarcity, consequently enhancing the overall network performance.
\citet{bansal2019pre} introduced a straightforward methodology to enhance direct speech-to-text translation (ST) in scenarios where the source language is low-resource. The approach involves initial pre-training of the model on a high-resource automatic speech recognition (ASR) task, followed by a subsequent fine-tuning process to refine its parameters specifically for speech-to-text translation (ST).
\citet{zhang2021pdaln} proposed an innovative approach, the adaptive data augmentation fine-tuning technique, designed to facilitate the efficient transfer of Named Entity Recognition (NER) knowledge from resource-rich domains to low-resource target domains.
\citet{liu2022improved} utilized pre-training data in both the initial pre-training and subsequent fine-tuning stages, strategically enhancing the model's performance across low-resource datasets. 
This dual-stage utilization of pre-training data contributes to a comprehensive and effective optimization, addressing the challenges posed by limited data availability in low-resource scenarios.
\citet{10103146} investigated the prospect of enhancing the performance of low-resource non-English languages by incorporating pre-trained language models that are primarily trained on English. 
This exploration seeks to leverage the knowledge embedded in English-dominant language models to boost the capabilities of models applied to non-English languages with limited resources.
To generate high-quality definition for low-resource language, \citet{zhang-etal-2023-assisting} leveraged a multilingual pre-trained model as backbone and employed a prompt contrastive fine-tuning approach to enhance the model's capabilities in this specific linguistic context.

In this paper, we adhere to the "pre-training and fine-tuning" paradigm, opting for a strategic approach rather than direct training of a DLKT model on low-resource KT datasets. 
This choice is made to mitigate potential overfitting issues and enhance the overall performance of model in low-resource KT datasets scenarios.

%% file: problem_statement.tex
The objective of KT problem is to predict the probability of whether a student will answer arbitrary $q_*$ correctly based on the student’s historical interaction data.
Specifically, suppose a student's chronologically ordered collection of $T$ past interactions is denoted as $\mathbf{S} = \{\mathbf{s}_j\}_{j=1}^T$, each student interaction $\mathbf{s}_j$ is represented as an ordering 4-tuple, i.e., $\mathbf{s}_j = <q_j, \{c | c \in \mathcal{N}_{q_j} \}_j, r_j, t_j>$, where $q_j$, $\{c\}_j$, $r_j$ and $t_j$ represent the specific question, the associated KC set, student response\footnote{Response $r_j \in \{0,1\}$, 1 represents the student answered correctly, and 0 otherwise.} and student's response timestamp respectively. 
$\mathcal{N}_{q_j}$ is the set of KCs that are associated with the question $q_j$. 
We would like to estimate the probability $\hat{r}_{*}$ of the student's future performance on arbitrary question $q_*$.

%% file: method.tex
In this section, we introduce the procedures in our proposed \emph{LoReKT} framework in details: 
(1) obtaining a foundational pre-trained KT model through learning from rich-resource KT datasets during the pre-training stage (Section \ref{sec:pretrain});
(2) efficiently adapting the pre-trained KT model to the low-resource dataset using an importance vector in the fine-tuning stage (Section~\ref{sec:finetune}).
\begin{figure*}[!thbp]
    \centering
    \includegraphics[width=1\columnwidth]{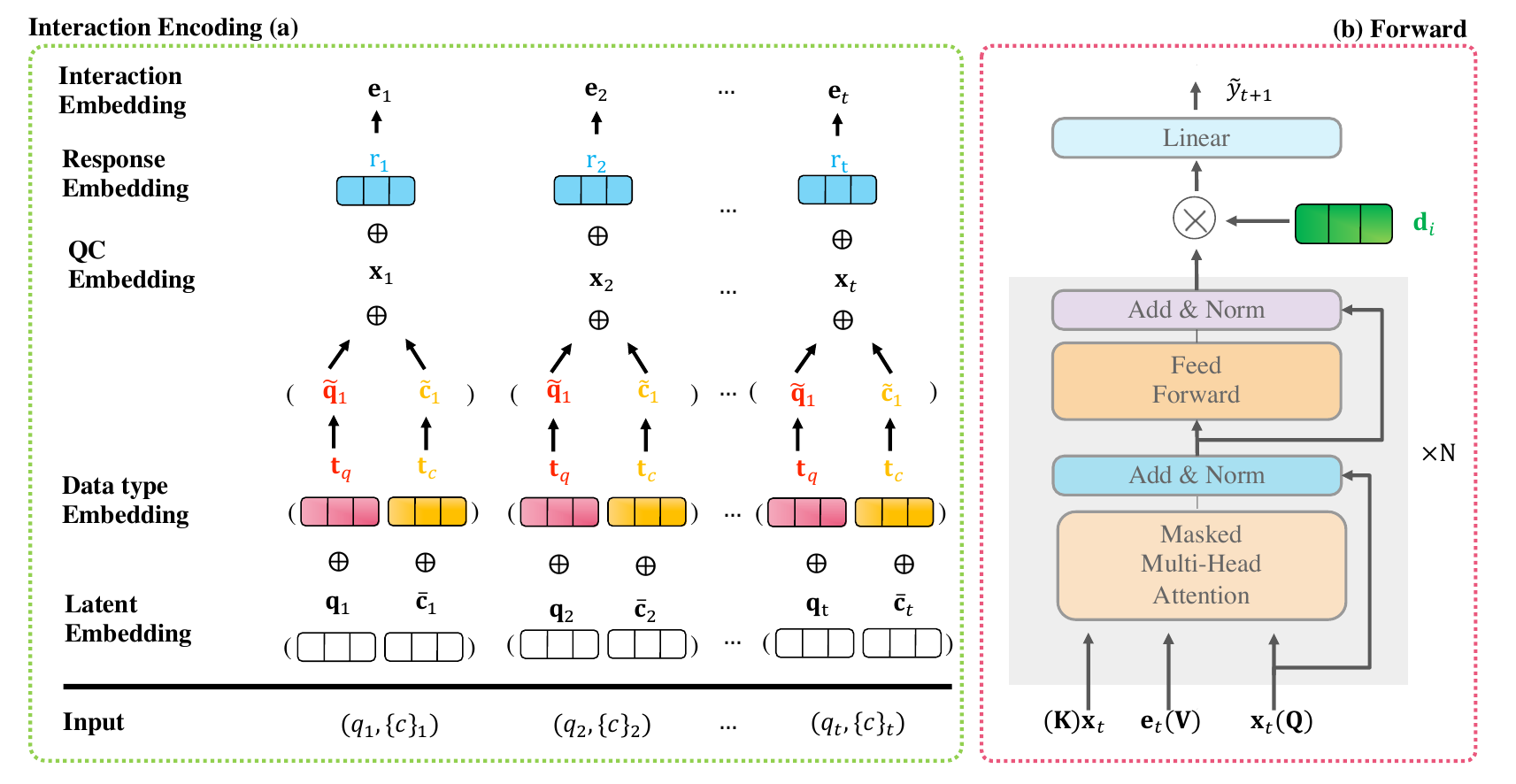}
    \caption{An illustration of the interaction encoding and forward procedure of our \emph{LoReKT} framework.}
    \label{fig:overview}
\end{figure*}

\subsection{Pre-training Stage}
\label{sec:pretrain}
\input{pretrain.tex}

\begin{figure*}[!thbp]
    \centering
    \includegraphics[width=1\columnwidth]{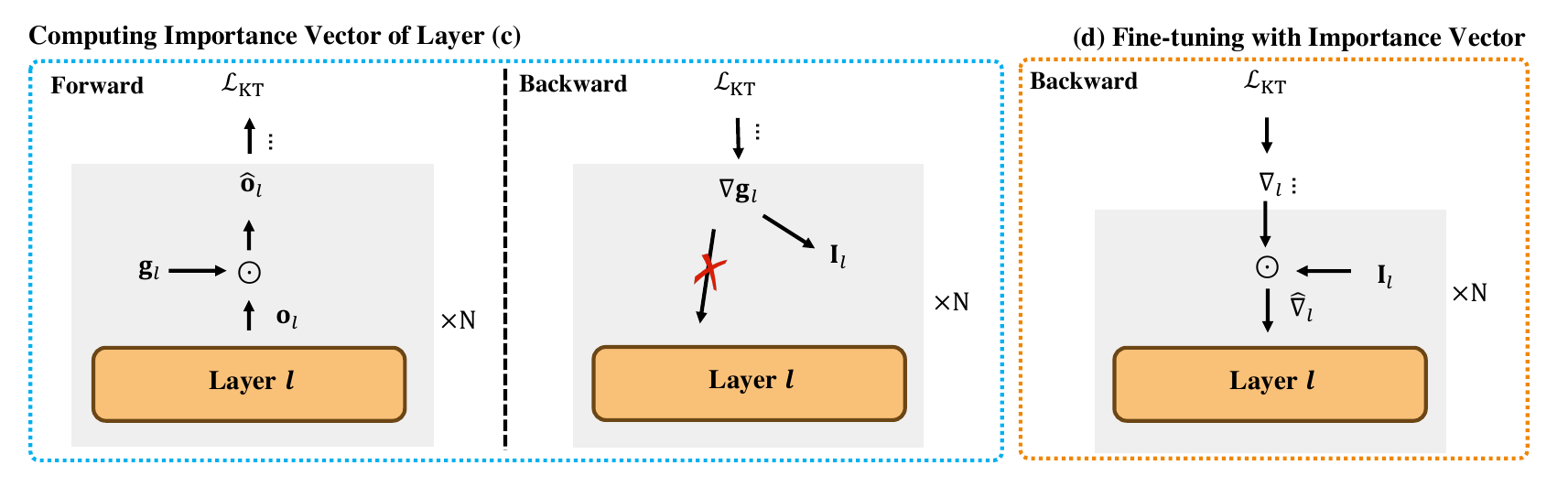}
    \caption{An illustration of the importance vector computing and applying procedure in our \emph{LoReKT} framework.}
    \label{fig:fine-tune}
\end{figure*}
\subsection{Fine-tuning Stage}
\label{sec:finetune}
\input{finetune.tex}

%% file: pretrain.tex
Our objective in the pre-training stage is to learn transferable parameters and representations from certain rich-resource datasets and build a pre-trained KT foundation model that is able to quickly adapt to low-resource KT datasets.

\subsubsection{Interaction Encoding}
\label{sec:interaction-encoding}
Due to the fact that the question bank is typically much larger than KCs, previous research mainly used KCs for interaction encoding, treating questions with the same KCs as identical \citep{piech2015deep,dkt+,dkt-forget}.
However, this approach led DLKT models to overlook the unique characteristics of same-KC questions, limiting interaction representation.
To address this, we align with works that include both individual question features and KCs to encode interactions in a more granular manner \citep{choi2020towards,ghosh2020context,liu2023simplekt}.

Specifically, let $\mathbf{D} = \{D_i\}_{i=1}^I $ be the mixed students' learning sequences, where $D_i = \{\mathbf{S}^{i}_{1}, ..., \mathbf{S}^{i}_{n}\}$. $\mathbf{S}^i_j$ is the $j$th learning sequence from rich-dataset $i$. $n$ is the number of learning sequences in $D_i$, and $I$ is the number of rich-resource KT datasets. Let $\mathcal{N}_{q_t}$ be the set of KCs associated with $q_t$. We represent question $q_t$ and its corresponding KCs as follows:

\begin{align}
    & \mathbf{q}_{t} = \mathbf{W}^q \cdot \mathbf{e}^q_t \nonumber \\
    & \bar{\mathbf{c}}_{t} = \frac{1}{\left|C_{q_t}\right|} \sum_{j = 1}^M \mathbf{c}_{j} * \mathbb{I}(c_j \in \mathcal{N}_{q_t}) \\
    & \mathbf{c}_{j} = \mathbf{W}^c \cdot \mathbf{e}^c_j  \nonumber
\end{align}

\noindent where $\mathbf{e}^q_t \in \mathbb{R}^{N \times 1}$ and $\mathbf{e}^c_j \in \mathbb{R}^{M \times 1}$ are the one-hot vectors that indicating the question $q_t$ and the related KC in $\mathcal{N}_{q_t}$.
$\mathbf{c}_{j} \in \mathbb{R}^{d \times 1}$ is one of the latent representations of the related KC to question $q_t$.
$\mathbf{q}_{t} \in \mathbb{R}^{d \times 1}$ and $\bar{\mathbf{c}}_{t} \in \mathbb{R}^{d \times 1}$ are the latent embedding of $q_t$ and its corresponding KCs, respectively.
$\mathbf{W}^q \in \mathbb{R}^{d \times N}$ and $\mathbf{W}^c \in \mathbb{R}^{d \times M}$ are learnable linear transformation operations.
$N$ and $M$ are the total number of distinct questions and KCs in our mixed dataset $\mathbf{D}$, respectively\footnote{We reassign ID numbers for all questions and KCs in the dataset $\mathbf{D}$ based on the values of $N$ and $M$. In the fine-tuning stage, for specific low-resource KT dataset, we adjust the ID numbers for its questions and KCs, starting from $N$ and $M$. Additionally, we expand the size of $\mathbf{W}^q$ and $\mathbf{W}^c$ to obtain their corresponding question and KC representations, i.e., $\mathbf{q}_t$ and $\mathbf{c}_{j}$.}.
$\cdot$ is the standard matrix/vector multiplication. 
$C_{q_t}$ is the size of $\mathcal{N}_{q_t}$ and $\mathbb{I}(\cdot)$ is the indicator function.

Drawing inspiration from the powerful pre-trained model BERT~\citep{bert}, which leverages token type embeddings to improve the integration of various token information, we introduce data type embeddings to the KT problem.
In this problem, there are two distinct data types: questions and concepts.
To improve the pre-trained KT model's ability to incorporate information from both, we introduce question and concept data type embeddings, which are directly integrated into all question and concept embeddings:

\begin{align}
    & \widetilde{\mathbf{q}}_{t} = \mathbf{q}_{t} \oplus \mathbf{t}_{q}; \quad \widetilde{\mathbf{c}}_t = \bar{\mathbf{c}}_{t} \oplus \mathbf{t}_{c} 
\end{align}

\noindent where $\mathbf{t}_{q} \in \mathbb{R}^{d \times 1}$ and $\mathbf{t}_{c} \in \mathbb{R}^{d \times 1}$ are the question and concept data type embeddings, $\oplus$ is the element-wise addition operator.
$\widetilde{\mathbf{q}}_{t}$ and $\widetilde{\mathbf{c}}_t$ are the question and KCs embeddings enriched with data type information.

Finally, we combine the embedding of question, its corresponding KCs, and response to encode the interaction $\mathbf{e}_t \in \mathbb{R}^{d \times 1}$, i.e.:

\begin{align}
    & \mathbf{x}_{t} = \widetilde{\mathbf{q}}_{t} \oplus \widetilde{\mathbf{c}}_t; \quad \mathbf{r}_t = \mathbf{W}^a \cdot \mathbf{a}^q_t; \quad \mathbf{e}_{t} = \mathbf{x}_{t} \oplus \mathbf{r}_{t} 
\end{align}

\noindent where $\mathbf{x}_t$ is the question-concept (QC) embedding, $\mathbf{a}^q_t \in \mathbb{R}^{2 \times 1}$ is the one-hot vector indicating whether the question $q_t$ is answered correctly and $\mathbf{W}^a \in \mathbb{R}^{d \times 2}$ is learnable linear transformation operation.
The illustration of the interaction encoding procedure is shown in Figure \ref{fig:overview} (a).

\subsubsection{Pre-training}
\label{sec:prediction}
Recently, generative pre-trained models that based on Transformer architecture have achieved promising results in various tasks compared to designing a sophisticated neural network for a specific task \citep{radford2018improving,brown2020language,gira2022debiasing}.
Drawing inspiration from these impressive findings, we opt to directly utilize a stack of transformer decoders. 
This choice enables us to dynamically capture student knowledge states without additional architecture design efforts:
\begin{align}
& \mathbf{h}^{(0)} = \mathbf{E} \oplus \mathbf{P} \nonumber \\
& \mathbf{h}^{(l)} = \mathbf{Tranformer\_block}(\mathbf{h}^{(l-1)}) \quad \forall{l}  \in [1,L] 
\end{align}

\noindent where $\mathbf{E} = (\mathbf{e}_{1}, ..., \mathbf{e}_{T})$ is the embedding matrix of $T$ past interaction. 
$\mathbf{P}$ is the position embedding matrix. 
$L$ is the number of layers. $\mathbf{h}^{(l)} \in \mathbb{R}^{T \times d}$ is a knowledge state embedding matrix of a student by $T$ past interactions.
Please note that to estimate student knowledge states via their historical interactions, we use QC embedding $\mathbf{x}_{t}$ for mapping both queries and keys, and interaction embedding $\mathbf{e}_{t}$ for mapping values in the self-attention mechanism.

Furthermore, inspired by prompt learning techniques that effectively capture diverse and overlapping patterns in multi-task learning scenarios \citep{raffel2020exploring,lester-etal-2021-power,sanh2021multitask,chung2022scaling}, we introduce dedicated dataset embeddings for individual KT datasets.
Given that each KT dataset has its unique prediction paradigm stemming from variations in question banks and KCs, this enhancement empowers the pre-trained KT model to effectively capture the specific and shared information across different KT datasets.
Specifically, the knowledge state $\mathbf{h}^{(l)}$ is first concatenated with corresponding dataset embedding $\mathbf{d}_i$ and QC embedding $\mathbf{x}_{t+1}$, then fed into a two-layer fully connected network with Sigmoid activation function $\sigma(\cdot)$ to predict the performance of a student on next question $q_{t+1}$:

\begin{align}
    & \mathbf{d}_{i} = \mathbf{W}^d \cdot \mathbf{e}^d \nonumber \\
    & \mathbf{y}_{t+1} = \mbox{ReLU} ( \mathbf{W}_1 \cdot [\mathbf{h}_{t+1}^{(l)}; \mathbf{x}_{t+1}; \mathbf{d}_{i}] + \mathbf{b}_1) \\
    & \hat{r}_{t+1} = \sigma (\mathbf{w}^\top \cdot \mbox{ReLU} \bigl( \mathbf{W}_2 \cdot \mathbf{y}_{t+1} + \mathbf{b}_2 \bigl) + b) \nonumber
    \label{eq:hidden_state2}
\end{align}

\noindent where $\mathbf{W}^d \in \mathbb{R}^{d \times I}$ is learnable linear transformation operation\footnote{In the fine-tuning stage, we expand the size of $\mathbf{W}^d$ to assign the dataset embedding for the specific low-resource KT dataset.}, $I$ is the total number of rich-resource KT datasets in the pre-training stage.
$\mathbf{e}^d \in \mathbb{R}^{I \times 1}$ is the one-hot vector indicating the corresponding dataset that current interaction belongs to.
$\mathbf{W}_1 \in \mathbb{R}^{d \times 2d}$, $\mathbf{W}_2 \in \mathbb{R}^{d \times d}$, $\mathbf{w} \in \mathbb{R}^{d \times 1}$, $\mathbf{b}_1 \in \mathbb{R}^{d \times 1}$, $\mathbf{b}_2 \in \mathbb{R}^{d \times 1}$ and $b$ are trainable parameters.
All learnable parameters in \emph{LoReKT} are trained in end-to-end fashion by minimizing the binary cross entropy loss between predicted probability $\hat{r}_{t}$ and the ground-truth label $r_t$:

\begin{equation}
 \mathcal{L}_{\text{KT}} = - \sum_{t=1}^{T} \bigl( r_t \log \hat{r}_t + (1-r_t) \log (1-\hat{r}_t) \bigl)
\label{eq:kt_loss}
\end{equation}

\noindent The forward procedure is illustrated in Figure~\ref{fig:overview} (b).

%% file: finetune.tex
In low-resource scenarios, overfitting is a common problem, as some model parameters may learn and memorize noisy dataset information, thereby hindering the model's ability to generalize.
This problem is especially severe in the low-resource KT setting. 
To alleviate the above problem, we introduce a novel importance vector-based fine-tuning strategy to encourage model to focus on updating the important parameters while constraining less important ones.

\subsubsection{Computing Importance Vector of Layer}
\label{sec:compute_imp_vector}
The backbone of \emph{LoReKT} is a stack of transformer decoders.
The key components of a transformer decoder are multi-head attention layer, intermediate layer, and output layer\footnote{In this paper, we use ``layer'' or $l$ to indicate any of these three layers, because the procedure of computing these three layers' importance vector is similar.}. 
It has been found that not all units (neurons or attention heads) in a specific layer are important \citep{michel2019sixteen}.
Therefore, before directly fine-tuning the model on each low-resource KT dataset, we adopt the approach described by \citet{ke2022continual} to compute the importance vector for each layer.
This is achieved by employing a gradient-based importance detection method, which is specifically tailored for each low-resource KT dataset:

\vspace{-0.5cm}
\begin{align}
& \hat{\mathbf{o}}_l = \mathbf{g}_l \odot \mathbf{o}_l \nonumber \\
& \mathbf{I}_l = \frac{1}{N} \sum_{n=1}^N \left| \frac{\partial \mathcal{L}_{\text{KT}}}{\partial \mathbf{g}_l} \right|
\label{eq:imp_vector}
\end{align}

\noindent where $\mathbf{o}_l$ refers to the output of layer $l$ (which can be any of the three layers mentioned above). 
$\odot$ refers to element-wise multiplication.
$\mathbf{g}_l$ serves as a \emph{virtual parameter}, sharing the same dimensions as $\mathbf{o}_l$, with each of its elements initialized to 1.
It remains unchanged during the computing process, as we only need its gradient on each parameter to get the importance of corresponding unit.
The unit with a higher gradient value obtained by its virtual parameter is considered more important, as they have a significant impact on the loss.
Therefore, the gradient of each parameter $g_{l,j}$ in $\mathbf{g}_l$ can be regarded as the importance of unit $j$ in layer $l$.
$\mathbf{I}_l$ is the importance vector of layer $l$, which is of the same size as ${\mathbf{g}_l}$, $N$ is the number of samples in current low-resource KT dataset and the $\mathcal{L}_{\text{KT}}$ is the loss defined in Eq.(\ref{eq:kt_loss}).
Noted that, each low-resource KT dataset has its own $\mathbf{I}_l$ for layer $l$.
The $\mathcal{L}_{\text{KT}}$ loss for each low-resource KT dataset is computed based on the zero-shot performance of pre-trained KT model (as obtained in Section \ref{sec:pretrain}).
The $\mathbf{g}_l$ remains unchanged during the computing process, because we need only its average gradient $\nabla \mathbf{g}_l$ (the term within $\left| \right|$ in eq.(\ref{eq:imp_vector})) over all the samples in the low-resource KT dataset and will not use the gradient to update the $\mathbf{g}_l$.
The illustration of computing importance vector of layer $l$ is shown in Figure~\ref{fig:fine-tune} (c).

\subsubsection{Fine-tune with Importance Vector}
\label{sec:finetune_with_impt}
After obtaining the importance vector $\mathbf{I}_l$ for each layer in each low-resource KT dataset using the pre-trained model, we initially compute the original gradient $\nabla_l$ by employing $\mathcal{L}_{\text{KT}}$ defined in Eq.~\ref{eq:kt_loss}.
Subsequently, we apply the importance vector $\mathbf{I}_l$ to obtain the modified gradient $\hat{\nabla}_l$ for updating:
\begin{align}
    & \hat{\nabla}_l = \mathbf{I}_l \odot \nabla_l \nonumber
    \label{eq:gradient_change}
\end{align}

\noindent Here, we expand (by copying) the $\mathbf{I}_l$ to match the dimensions of $\nabla_l$ to apply it to all associated parameters.
The modified gradient $\hat{\nabla}_l$ is only employed in the backward pass.
This encourages the model to prioritize updating the associated parameters with high importance instead of less important ones by regulating their gradient flow.
The procedure of fine-tuning based on importance vector is shown in Figure~\ref{fig:fine-tune} (d).

%% file: exp.tex
\begin{table*}[t]
    \fontsize{10}{10}\selectfont
    \setlength\tabcolsep{4pt}
    \centering
    \begin{tabular}{lccccccc}
    \toprule
                       & \multicolumn{3}{c}{\textbf{Low-resource}}  &  & \multicolumn{3}{c}{\textbf{Rich-resource}}  \\ \cline{2-4} \cline{6-8} \addlinespace[2pt]
                       & AS2009  & NIPS34    & AL2005  & & BD2006    & XES3G5M    & EdNet       \\ \midrule
    \# of Ques.        & 26,688  & 948       & 210,710 & & 207,856   & 7,652     & 12,235      \\
    \# of KCs          & 123     & 57        & 112     & & 493       & 865       & 188         \\
    \# of Interactions & 346,860 & 1,382,727 & 809,694 & & 3,679,199 & 5,549,635 & 6,533,522   \\
    avg KCs            & 1.1969  & 1.0148    & 1.3634  & & 1.0136    & 1.1640    & 2.2611      \\
    Subject            & Math    & Math      & Math    & & Math      & Math      & Linguistics \\
    Language           & English & English   & English & & English   & Chinese   & English     \\ 
    \bottomrule
    \end{tabular}
    \caption{Dataset statistics of 6 datasets. “avg KCs” denotes the number of average KCs per question.}
    \label{tab:dataset}
\end{table*}

In this section, we present details of our experiment settings and the corresponding results. 
We conduct comprehensive analysis to illustrate the effectiveness of our \emph{LoReKT} framework. 
Specifically, we aim to answer the following research questions: 
(\textbf{RQ1}) Can we build a solid pre-trained foundational model for KT? 
(\textbf{RQ2}) In low-resource scenarios, how does our proposed \emph{LoReKT} framework performs compared to the state-of-the-art KT methods?
(\textbf{RQ3}) Does pre-training truly enhance the performance of mode in low-resource KT datasets?
(\textbf{RQ4}) In the fine-tuning stage, is it effective to focus on updating important parameters based on $\mathbf{I}_l$?
(\textbf{RQ5}) How does the dataset and data type embedding affect the pre-trained KT model?

\subsection{Datasets}
\input{data.tex}

\subsection{Experimental Setting}
\label{sec:exp_setting}
\input{exp_setting.tex}

\subsection{Baselines}
\label{sec:baselines}
\input{baselines.tex}

\begin{table}[!b]
    \fontsize{8}{11}\selectfont
    \centering
    \setlength\tabcolsep{1pt}
    \begin{tabular}{lccccccc}
    \toprule
    \multicolumn{1}{l}{\multirow{2}{*}{\textbf{Method (Chronologically)}}} & \multirow{2}{*}{\textbf{Model Type}}  & \multicolumn{6}{c}{\textbf{AUC}}                                                                         \\ \cline{3-8} \addlinespace[2pt]
    \multicolumn{1}{c}{}                     &                                                                                    & \textbf{AS2009} & \textbf{NIPS34} & \textbf{AL2005} & \textbf{BD2006} & \textbf{XES3G5M} & \textbf{EdNet} \\ \hline
    DKT \citep{piech2015deep}               & Sequential                                                                                      & 0.7525          & 0.7688          & 0.8159          & 0.8018          & 0.7845          & 0.6405         \\
    DKVMN  \citep{zhang2017dynamic}         & Memory                                                                                          & 0.7472          & 0.7677          & 0.8052          & 0.7999          & 0.7796          & 0.6576         \\
    DKT+ \citep{yeung2018addressing}        & Sequential                                                                                      & 0.7543          & 0.7698          & 0.8141          & 0.8019          & 0.7858          & 0.6454         \\
    DKT-F \citep{nagatani2019augmenting}    & Sequential                                                                                      & -               & 0.7728          & 0.8146          & 0.7997          & 0.7935          & 0.6548         \\
    KQN \citep{lee2019knowledge}            & Sequential                                                                                      & 0.7462          & 0.7685          & 0.8010          & 0.7953          & 0.7794          & 0.6415         \\
    SKVMN  \citep{abdelrahman2019knowledge} & Memory                                                                                          & 0.7332          & 0.7513          & 0.7463          & 0.7287          & 0.7512          & 0.6374         \\
    DeepIRT  \citep{yeung2019deep}          & Memory                                                                                          & 0.7465          & 0.7673          & 0.8040          & 0.7976          & 0.7789          & 0.6387         \\
    GKT \citep{nakagawa2019graph}           & Others                                                                                           & 0.7442          & 0.7718          & 0.8112          & 0.8041          & 0.7731          & 0.6392         \\
    SAKT \citep{pandey2019self}             & Attention                                                                                       & 0.7221          & 0.7508          & 0.7850          & 0.7748          & 0.7685          & 0.6290         \\
    SAINT \citep{choi2020towards}           & Attention                                                                                       & 0.6990          & 0.7883          & 0.7764          & 0.7758          & 0.8070          & 0.6841         \\
    AKT \citep{ghosh2020context}            & Attention                                                                                       & \underline{0.7869}          & \textbf{0.8038}          & 0.8324          & \textbf{0.8213}          & \textbf{0.8215}          & 0.7054         \\
    ATKT \citep{guo2021enhancing}           & Others                                                                                     & 0.7472          & 0.7664          & 0.7987          & 0.7889          & 0.7791          & 0.6490         \\
    HawkesKT  \citep{wang2021temporal}      & Others                                                                                          & 0.7232          & 0.7763          & 0.8199          & 0.8077          & 0.7933          & 0.7304         \\
    LPKT \citep{shen2021learning}           & Sequential                                                                                      & 0.7812          & 0.8004          & 0.8268          & 0.8056          & 0.8163          & 0.7644         \\
    AT-DKT \citep{liu2023enhancing}         & Sequential                                                                                      & 0.7555          & 0.7816          & 0.8246          & 0.8104          & 0.7925          & 0.6536         \\
    simpleKT \citep{liu2023simplekt}        & Attention                                                                                       & 0.7744          & \underline{0.8035}          & 0.8254          & 0.8160          & 0.8161          & 0.6765         \\
    sparseKT \citep{huang2023towards}       & Attention                                                                                       & 0.7739          & 0.8033          & 0.8152          & 0.8120          & 0.8165          & 0.6804         \\ \hline \addlinespace[2pt]
    \emph{LoReKT-Base-89M}                        & Attention                                                                                    & 0.6041          & 0.6401          & 0.5943          & 0.8049          & 0.8145          & \underline{0.7647}         \\ 
    \emph{LoReKT-Base-221M}                       & Attention                                                                                    & 0.6228          & 0.6452          & 0.6155          & \underline{0.8183}          & \underline{0.8192}          & \textbf{0.7672}         \\
    \emph{LoReKT-Base-478M}                       & Attention                                                                                    & 0.5957          & 0.6103          & 0.5834          & 0.8061          & 0.8164          & 0.7659         \\
    \emph{LoReKT-Base-1.01B}                      & Attention                                                                                    & 0.5761          & 0.5980          & 0.5745          & 0.8003          & 0.8121          & 0.7633         \\ \hline \addlinespace[2pt]
    \emph{LoReKT-Ft-impt-221M}                       & Attention                                                                                 & \textbf{0.7912}          & 0.8002          & \textbf{0.8425}          & -               & -               & -         \\
    \emph{LoReKT-Ft-221M}                       & Attention                                                                                    & 0.7833          & 0.7969          & \underline{0.8359}          & -               & -               & -         \\
    \bottomrule
    \end{tabular}
    \caption{The overall performance in terms of AUC. The result of each low-resource KT dataset corresponds to a separately fine-tuned model, leading to different performance on pre-training datasets. Therefore, we use ``-'' to denote the results on pre-training datasets. The best result is indicated in bold, while the second best result is denoted in \underline{underline}.}
    \label{tab:overall_auc}
\end{table}
\begin{table}[!t]
    \fontsize{8}{11}\selectfont
    \centering
    \setlength\tabcolsep{1pt}
    \begin{tabular}{lccccccc}
    \toprule
    \multicolumn{1}{l}{\multirow{2}{*}{\textbf{Method (Chronologically)}}} & \multirow{2}{*}{\textbf{Model Type}}  & \multicolumn{6}{c}{\textbf{Accuracy}}                                                                         \\ \cline{3-8} \addlinespace[2pt]
    \multicolumn{1}{c}{}                     &                                                                                    & \textbf{AS2009} & \textbf{NIPS34} & \textbf{AL2005} & \textbf{BD2006} & \textbf{XES3G5M} & \textbf{EdNet} \\ \hline
    DKT \citep{piech2015deep}               & Sequential                                                                                  & 0.7228          & 0.7031          & 0.8105          & 0.8554          & 0.8173          & 0.6665         \\
    DKVMN  \citep{zhang2017dynamic}         & Memory                                                                                      & 0.7196          & 0.7022          & 0.8026          & 0.8547          & 0.8155          & 0.6392         \\
    DKT+ \citep{yeung2018addressing}        & Sequential                                                                                  & 0.7243          & 0.7046          & 0.8085          & 0.8554          & 0.8179          & 0.6668         \\
    DKT-F \citep{nagatani2019augmenting}    & Sequential                                                                                  & -               & 0.7076          & 0.8092          & 0.8540          & 0.8209          & 0.6666         \\
    KQN \citep{lee2019knowledge}            & Sequential                                                                                  & 0.7214          & 0.7028          & 0.8022          & 0.8540          & 0.8154          & 0.6665         \\
    SKVMN  \citep{abdelrahman2019knowledge} & Memory                                                                                     & 0.7156          & 0.6885          & 0.7837          & 0.8406          & 0.8071          & 0.6572         \\
    DeepIRT  \citep{yeung2019deep}          & Memory                                                                                     & 0.7196          & 0.7020          & 0.8029          & 0.8542          & 0.8152          & 0.6559         \\
    GKT \citep{nakagawa2019graph}           & Others                                                                                      & 0.7179          & 0.7053          & 0.8078          & 0.8555          & 0.8139          & 0.6672         \\
    SAKT \citep{pandey2019self}             & Attention                                                                                   & 0.7031          & 0.6873          & 0.7948          & 0.8460          & 0.8121          & 0.6519         \\
    SAINT \citep{choi2020towards}           & Attention                                                                                   & 0.6977          & 0.7187          & 0.7770          & 0.8455          & 0.8177          & 0.6624         \\
    AKT \citep{ghosh2020context}            & Attention                                                                                   & \underline{0.7385}          & 0.7320          & 0.8138          & \underline{0.8594}          & \textbf{0.8275}          & 0.6888         \\
    ATKT \citep{guo2021enhancing}           & Others                                                                                      & 0.7206          & 0.7012          & 0.7989          & 0.8510          & 0.8143          & 0.6642         \\
    HawkesKT  \citep{wang2021temporal}      & Others                                                                                     & 0.7046          & 0.7110          & 0.8108          & 0.8563          & 0.8191          & 0.7076         \\
    LPKT \citep{shen2021learning}           & Sequential                                                                                  & 0.7355          & 0.7309          & \underline{0.8154}          & 0.8547          & 0.8264          & 0.7243         \\
    AT-DKT \citep{liu2023enhancing}         & Sequential                                                                                  & 0.7250          & 0.7146          & 0.8144          & 0.8560          & 0.8195          & 0.6684         \\
    simpleKT \citep{liu2023simplekt}        & Attention                                                                                   & 0.7320          & \textbf{0.7328}          & 0.8083          & 0.8579          & 0.8240          & 0.6624         \\
    sparseKT \citep{huang2023towards}       & Attention                                                                                   & 0.7282          & 0.7322          & 0.8017          & 0.8569          & 0.8234          & 0.6643         \\ \hline
    \emph{LoReKT-Base-89M}                        & Attention                                                                                & 0.6035          & 0.6003          & 0.6753          & 0.8543          & 0.8248          & \underline{0.7243}         \\ 
    \emph{LoReKT-Base-221M}                       & Attention                                                                                & 0.6216          & 0.6008          & 0.6962          & \textbf{0.8596}          & \underline{0.8271}          & \textbf{0.7250}         \\
    \emph{LoReKT-Base-478M}                       & Attention                                                                                & 0.5929          & 0.5821          & 0.6644          & 0.8538          & 0.8253          & 0.7212         \\
    \emph{LoReKT-Base-1.01B}                      & Attention                                                                                & 0.5805          & 0.5745          & 0.6568          & 0.8501          & 0.8239          & 0.7207         \\ \hline \addlinespace[2pt]
    \emph{LoReKT-Ft-impt-221M}                       & Attention                                                                               & \textbf{0.7402}          & \underline{0.7323}          & \textbf{0.8242}          & -               & -               & -         \\
    \emph{LoReKT-Ft-221M}                       & Attention                                                                                & 0.7353          & 0.7275          & 0.8159          & -               & -               & -         \\
    \bottomrule
    \end{tabular}
    \caption{The overall performance in terms of Accuracy. The result of each low-resource KT dataset corresponds to a separately fine-tuned model, leading to different performance on pre-training datasets. Therefore, we use ``-'' to denote the results on pre-training datasets. The best result is indicated in bold, while the second best result is denoted in \underline{underline}.}
    \label{tab:overall_acc}
\end{table}

\subsection{Results}
We utilize different variants of \emph{LoReKT} to represent its performance under different settings.
\emph{LoReKT-Base} represents the model trained after the pre-training stage without any fine-tuning on a specific low-resource dataset. 
\emph{LoReKT-Ft-impt} and \emph{LoReKT-Ft} refer to the model that fine-tuned on the specific low-resource KT dataset with and without using importance vector. 
\subsubsection{Model Performance after Pre-training (RQ1)}

\label{sec:pretrain_result}
\input{pretrain_result.tex}

\subsubsection{Fine-tuning Performance of LoReKT (RQ2)}
\label{sec:finetune_result}
\input{finetune_result.tex}

\subsubsection{Impact of Pre-training (RQ3)}
\label{sec:abl_pretrain}
\input{abl_pretrain.tex}

\begin{figure}[!thbp]
    \centering
    \includegraphics[width=0.7\columnwidth]{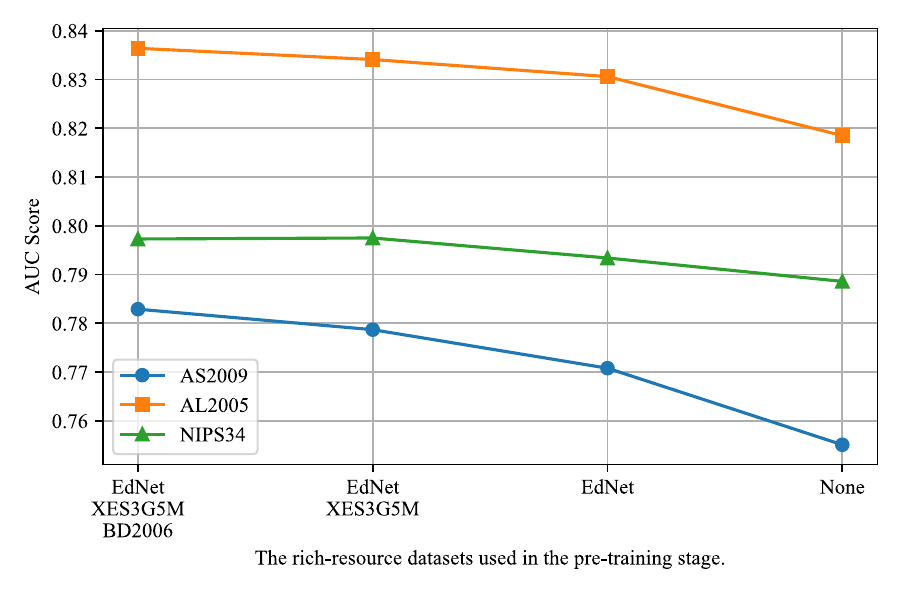}
    \caption{The impact of the number of rich-resource KT datasets used in the pre-training stage on the performance of low-resource KT datasets.}
    \label{fig:pretrain_num}
\end{figure}

\subsubsection{Impact of Importance Vector (RQ4)}
\label{sec:abl_impt}
\input{abl_impt.tex}

\begin{table}[]
    \fontsize{8.5}{11}\selectfont
    \centering
    \setlength\tabcolsep{1.4pt}
    \begin{tabular}{lccccccc}
    \toprule
    \multicolumn{1}{l}{\multirow{2}{*}{\textbf{Method}}}   & \multicolumn{3}{c}{\textbf{AUC}} & &  \multicolumn{3}{c}{\textbf{Accuracy}}     \\ \cline{2-4} \cline{6-8} \addlinespace[2pt]
    \multicolumn{1}{c}{}                                   & \textbf{BD2006} & \textbf{XES3G5M} & \textbf{EdNet} &  & \textbf{BD2006} & \textbf{XES3G5M} & \textbf{EdNet} \\ \hline \addlinespace[2pt]
    \emph{LoReKT-Base-221M}                                   & \textbf{0.8183}          & \textbf{0.8192}          & \textbf{0.7672}   &  & \textbf{0.8596}          & \textbf{0.8271}          & \textbf{0.7250}       \\ \hline \addlinespace[2pt]
    w/o data type \& dataset embedding                      & 0.8145          & 0.8139          & 0.7594   &  & 0.8541          & 0.8239          & 0.7185    \\
    w/o data type embedding                                & 0.8161          & \underline{0.8172}          & \underline{0.7646}   &   & \underline{0.8575}          & \underline{0.8268}          & 0.7243        \\
    w/o dataset embedding                                  & \underline{0.8173}         & 0.8168         & 0.7625  &  & 0.8569         & 0.8268         & \underline{0.7246}        \\
    \bottomrule
    \end{tabular}
    \caption{The impact of data type and dataset embedding in \emph{LoReKT-Base} in terms of AUC and Accuracy performance. ''w/o`` means excluding such module from \emph{LoReKT-Base}. The best result is indicated in bold, while the second best result is denoted in \underline{underline}.}
    \label{tab:abl_pretrain}
\end{table}
\subsubsection{Impact of Data Type and Dataset Embeddings (RQ5)}
\label{sec:abl_datatype}
\input{abl_datatype.tex}

%% file: data.tex
Since our \emph{LoReKT} framework learns transferable parameters and representations from rich-resource KT datasets first and then quickly adapt to low-resource scenarios, we select three rich-resource datasets including BD2006~\citep{algebra}, XES3G5M~\citep{liu2023xes3g5m}, and EdNet~\citep{ednet} to establish a robust foundational pre-trained KT model. 
We further fine-tune the pre-trained KT model on three low-resource KT datasets including AS2009~\citep{as2009}, NIPS34~\citep{nips}, and AL2005~\citep{algebra} respectively. 
The data statistics for the six selected datasets can be found in Table \ref{tab:dataset}.
The detailed descriptions are as follows:

\paragraph{Rich-resource KT Datasets}
\begin{itemize}
\item Bridge2algebra2006 (BD2006)\footnote{https://pslcdatashop.web.cmu.edu/KDDCup/}: this dataset is provided from the KDD Cup 2010 EDM Challenge with the algebra questions answered by 13-14 years old students.
\item XES3G5M\footnote{https://github.com/ai4ed/XES3G5M}: this large-scale dataset from a Chinese online mathematics learning platform includes rich information about student learning interactions.
\item EdNet\footnote{https://github.com/riiid/ednet}: this dataset from South Korea's Santa AI tutoring system is one of the largest KT datasets, with 130+ million student interactions in the TOEIC test.
\end{itemize}

\paragraph{Low-resource KT Datasets}
\begin{itemize}
\item ASSISTments2009 (AS2009)\footnote{https://sites.google.com/site/assistmentsdata/home/2009-2010-assistment-data/skill-builder-data-2009-2010}: this dataset is one of the classical education dataset that collects students' responses to mathematic questions from the free online tutoring ASSISTments platform during the school year 2009-2010.
\item NIPS34\footnote{https://eedi.com/projects/neurips-education-challenge}: this dataset is released in the NeurIPS 2020 Education Challenge. In our work, we choose to use Task 3\&4 that are the students' responses to multiple-choice diagnostic math questions on the Eedi platform.
\item Algebra2005 (AL2005)\footnote{https://pslcdatashop.web.cmu.edu/KDDCup/}: similar to BD2006, this dataset is also released on KDD Cup 2010 EDM Challenge. 
\end{itemize}

%% file: exp_setting.tex
We remove student sequences shorter than 3 attempts and truncate student interaction sequences that are longer than 200. We use 80\% of student sequences for training and validation and the rest 20\% of student sequences for model evaluation.
We adopt Adam optimizer \citep{kingma2015adam} to train all the models. 
The number of training epochs is set to 200.
Following all existing DLKT research \citep{liu2022pykt,liu2023simplekt,choi2020towards}, we use the Area Under the Curve (AUC) as the main evaluation metric and use Accuracy as the secondary evaluation metric.
We use an early stopping strategy that stops optimization when the AUC score fails to get the improvement on the validation set in the latest 10 epochs. 
Owing to the training efficiency, it is difficult to tune too many hyperparameters for the models with billions of parameters. 
Hence, we only tune the model learning rate in \{0.001, 0.0001\} with the dropout rate in \{0.1, 0.2\} for fair comparison in various model sizes.

With the aim of finding a suitable model size for pre-training, we conduct an extensive exploration of various model sizes, encompassing a range from 89M to 1.01B.
The corresponding architecture details are summarized in Table~\ref{tab:model_size}.
$n_{params}$ is the total number of trainable parameters. 
$n_{layers}$ is the total number of layers, $d_{model}$ is the number of units in each bottleneck layer (the feed-forward layer is denoted as $d_{ff}$), and $n_{head}$ is the number of attention heads.

\begin{table}[]
    \fontsize{10}{12}\selectfont
    \setlength\tabcolsep{4pt}
    \centering
    \begin{tabular}{lcccccc}
    \toprule
    Model Name & \multicolumn{1}{l}{$n_{params}$} & \multicolumn{1}{l}{$n_{layers}$} & \multicolumn{1}{l}{$d_{model}$} & \multicolumn{1}{l}{$n_{head}$} & \multicolumn{1}{l}{$d_{ff}$} \\ \midrule
    \emph{LoReKT-Base-89M} & 89M   & 4  & 256  & 8  & 256  \\
    \emph{LoReKT-Base-221M} & 221M  & 24 & 512  & 16 & 1024 \\
    \emph{LoReKT-Base-478M} & 478M  & 24 & 1024 & 16 & 1024 \\
    \emph{LoReKT-Base-1.01B} & 1.01B & 32 & 1536 & 24 & 2560 \\
    \bottomrule
    \end{tabular}
    \caption{The model sizes and associated architecture details of \emph{LoReKT-Base}.} 
    \label{tab:model_size}
\end{table}

%% file: baselines.tex
To conduct a comprehensive evaluation of \emph{LoReKT}, we have compared it with 17 selected baseline methods.
We have carefully categorized these baseline methods into four categories:
\paragraph{Deep Sequential KT Models}
Deep sequential KT models use an auto-regressive framework to dynamically track students' knowledge states.
Representative deep sequential KT models include:
\begin{itemize}
    \item DKT \cite{piech2015deep}: it uses an LSTM layer to model students' learning processes. 
    \item DKT+ \cite{yeung2018addressing}: it improves the original DKT model by addressing the reconstruction and inconsistent issues.
    \item DKT-F \cite{nagatani2019augmenting}: it enhances original DKT by considering students' forgetting behaviors.
    \item KQN \cite{lee2019knowledge}: it is a recurrent neural network (RNN) based architecture that extracts the relation representations between students' learning abilities and KCs to predict their performance.
    \item LPKT \cite{shen2021learning}: it designs a learning cell to model the students' learning processes to estimate their knowledge states.
    \item AT-DKT \cite{liu2023enhancing}: it proposes two auxiliary learning tasks involving question tagging prediction task and individualized prior knowledge prediction task to improve the prediction performance of DKT.
\end{itemize}

\paragraph{Memory Augmented KT Models}
Memory augmented KT models employ memory networks to capture potential relevances between KCs and student knowledge states. 
Representative memory augmented KT models include:
\begin{itemize}
    \item DKVMN \cite{zhang2017dynamic}: it incorporates a static matrix to store the relationships among KCs and a dynamic matrix to track the student's knowledge state.
    \item SKVMN \cite{abdelrahman2019knowledge}: it is a combination of DKVMN and LSTM that uses a hop-LSTM layer to capture sequential dependencies of questions.
    \item DeepIRT \cite{yeung2019deep}: it incorporates DKVMN and item response theory to enhance the interpretability of the prediction output of DKVMN.
\end{itemize}

\paragraph{Attention based KT Models}
Attention based KT models capture dependencies between historical interactions and the next questions via the attention mechanism.
Representative attention based KT models include:
\begin{itemize}
    \item SAKT \cite{pandey2019self}: it uses self-attention to identify the relevance between historical interactions and KCs.
    \item SAINT \cite{choi2020towards}: it is a Transformer-based model for KT that encodes questions and responses in the encoder and decoder respectively.
    \item AKT \cite{ghosh2020context}: it leverages three self-attention modules to estimate the relevance between questions and historical interactions and explicitly models student's forgetting behavior via a monotonic attention mechanism. 
    \item simpleKT \cite{liu2023simplekt}: it explores the ordinary dot-product attention based KT models by capturing the individual differences among questions covering the same set of KCs 
    \item sparseKT \cite{huang2023towards}: it incorporates a k-selection module to only pick items with the highest attention scores to improve the robustness and generalization of the attention based DLKT approaches.
\end{itemize}

\paragraph{Other KT Models}
Other KT models that do not belong to the above categories:
\begin{itemize}
    \item HawkesKT \cite{wang2021temporal}: it utilizes the Hawkes process to model temporal cross-effects in student historical interactions.
    \item ATKT \cite{guo2021enhancing}: it performs adversarial perturbations into the student interaction sequence to enhance the generalization ability based on an attention-LSTM based KT model.
    \item GKT \cite{nakagawa2019graph}: it casts the knowledge structure as a graph and reformulates the KT task as a time series node-level classification problem in GNN. 
\end{itemize}

%% file: pretrain_result.tex
We report the results of main evaluation metric, i.e., AUC, in Table \ref{tab:overall_auc} and the results of secondary evaluation metric, i.e., Accuracy, in Table~\ref{tab:overall_acc}.
From Table \ref{tab:overall_auc},  we have the following observations: 
(1) in the comparison of various model sizes in \emph{LoReKT-Base},
\emph{LoReKT-Base-221M} exhibits the best performance. Initially, as the model size increases, the model's performance improves; however, it begins to decline when the model size becomes excessively large.
We argue that this phenomenon is indicative of the model first experiencing underfitting followed by overfitting;
(2) the \emph{LoReKT-Base-221M} demonstrates strong performance across all three pre-training datasets (BD2006, XES3G5M, and EdNet).
For example, it achieves the highest AUC score on the EdNet dataset, showcasing a substantial improvement of 12.59\% over sparseKT, 9.2\% over LPKT and 8.6\% over AKT.
It ranks second in terms of AUC on BD2006 and XES3G5M datasets and is on par with the best model AKT within a 0.5\% range of performance gap.
It's noteworthy that \emph{LoReKT-Base-221M}, utilizing only a single model, consistently achieves strong performance across all three datasets. 
In contrast, the AKT model is separately trained for each dataset, resulting in significant performance variations. 
For instance, while the AKT model performs well on BD2006 and XES3G5M, its performance on EdNet is notably lower, lagging behind \emph{LoReKT-Base-221M} by 8.1\%.
Furthermore, the architecture of \emph{LoReKT-Base-221M} is more concise than AKT which designs two extra modules upon the original RNN architecture;
and (3) the \emph{LoReKT-Base-221M} also demonstrates impressive zero-shot capabilities on previously unseen datasets (AS2009, NIPS34, and AL2005). 
For example, it achieves AUC scores of 0.6452 and 0.6228 in NIPS34 and AS2009 datasets, which is significantly outperforming random performance. 
In spite of the zero-shot performance is still far from usable, it veriﬁes our conjecture that \emph{LoReKT-Base} has good potential transferability between different disciplines via cross-source learning.
Furthermore, its robust zero-shot capabilities enhance the accuracy of the importance vector computed in Section~\ref{sec:compute_imp_vector} for each low-resource KT dataset in the fine-tuning stage.

%% file: finetune_result.tex
After obtaining the pre-trained KT model \emph{LoReKT-Base-221M}, we further fine-tune it on each low-resource KT dataset.
From Table~\ref{tab:overall_auc}, we can observe that 
(1) comparing \emph{LoReKT-Base-221M} and \emph{LoReKT-Ft-impt-221M}, the performance on all three low-resource KT datasets is significantly improved in terms of AUC score (e.g., an improvement of 26.1\% in AS2009, 24.3\% in NIPS34 and 37.3\% in AL2005);
(2) the \emph{LoReKT-Ft-impt-221M} outperforms all the baseline methods in AS2009 and AL2005 datasets in terms of AUC score.
More specifically, compared with ATKT, sparseKT, and simpleKT, it significantly improves the AUC score by 5.61\%, 3.47\%, and 2.16\% on AL2005 respectively.
Also, it outperforms the majority of baseline methods and closely matches the performance of the top models such as AKT and simpleKT on NIPS34, with performance gap of less than 0.5\%.
Notably, our \emph{LoReKT-Ft-impt-221M} is robust enough to consistently achieve strong performance across all three low-resource datasets without the need for additional architecture design, while AKT and simpleKT exhibit notable performance variations across these datasets.
Furthermore, it can be observed that the performance of \emph{LoReKT-Ft-impt-221M} on NIPS34 is slightly lower than on AS2009 and AL2005 datasets when compared to the baseline methods.
We attribute this discrepancy to the larger dataset size of NIPS34 relative to AS2009 and AL2005 (shown in Table~\ref{tab:dataset}), which suggests that the potential overfitting issue is not as prominent in NIPS34.

%% file: abl_pretrain.tex
To analyze the impact of pre-training on the performance of low-resource KT datasets, we progressively reduce the number of rich-resource datasets used in the pre-training. 
From Figure~\ref{fig:pretrain_num}, we have the following observations:
(1) As the number of rich-resource KT datasets used in the pre-training stage decreases, the model's performance on low-resource KT datasets correspondingly drops.
(2) When pre-training is omitted, the model's performance on low-resource KT datasets significantly deteriorates, particularly for the AS2009 and AL2005 datasets. 
For example, there is a decline in the AUC score by 3.6\% in AS2009 and 2.1\% in AL2005, whereas only a 1.1\% decline is observed in NIPS34. 
This could be attributed to the fact that the data quantity in AS2009 and AL2005 is much smaller than in NIPS34, leading to a more pronounced overfitting issue.

%% file: abl_impt.tex
We conduct experiments to further investigate the effectiveness of our proposed fine-tuning strategy with importance vector in low-resource KT datasets.
Comparing the results of \emph{LoReKT-Ft-221M} and \emph{LoReKT-Ft-impt-221M} in Table~\ref{tab:overall_auc}, we observe that the proposed fine-tuning strategy with importance vector enhances performance across all low-resource datasets.
It leads to AUC improvements of 0.66\% for the AL2005 dataset and 0.79\% for the AS2009 dataset, which is larger than the 0.33\% improvement observed for the NIPS34 dataset.
This observation suggests that the fine-tuning strategy with importance vector is more effective when the dataset size is insufficient (as shown in Table~\ref{tab:dataset}, NIPS34 exhibits a higher interaction count compared to AS2009 and AL2005). 
We believe that this effectiveness is attributed to the importance vector, which restricts the update of unimportant parameters, thereby mitigating the risk of overfitting.

%% file: abl_datatype.tex
We also analyze the impact of data type and dataset embedding in Section~\ref{sec:pretrain}.
We report the AUC and Accuracy results in Table~\ref{tab:abl_pretrain}.
As presented in Table~\ref{tab:abl_pretrain}, comparing the results of ``w/o data type \& dataset embedding'' and ``w/o data type embedding'', we demonstrate that the proposed dataset embedding enhances the model's understanding of the corresponding dataset's prediction paradigm.
This effect is particularly pronounced for datasets with larger quantities, such as the improvement in AUC of 0.52\% for EdNet, surpassing the improvements of 0.33\% in XES3G5M and 0.16\% in BD2006.
We attribute this to the fact that the dataset embedding is well-represented when the corresponding dataset has a sufficient quantity (as indicated in Table~\ref{tab:dataset}).
The results of ``w/o data type \& dataset embedding'' and ``w/o dataset embedding'' reveal that, across all three datasets, the proposed data type embedding plays an important role in enabling the model to incorporate information from both questions and concepts (e.g., an improvement in AUC of 0.28\%, 0.29\% and 0.31\% in BD2006, XES3G5M, and EdNet respectively).

%% file: conclusion.tex
In this paper, we focus on improving the performance of DLKT models on low-resource KT datasets.
To address this problem, we propose a framework called \emph{LoReKT} based on a stack of transformer decoders.
The \emph{LoReKT} includes two stages: pre-training and fine-tuning, and does not require sophisticated architecture for a specific dataset.
In the pre-training stage, we establish a robust pre-trained KT model based on several rich-resource KT datasets.
Subsequently, we leverage an importance mechanism fine-tuning strategy to adapt the pre-trained model to a specific low-resource KT dataset effectively.
The extensive quantitative and qualitative experiment results on six real-world datasets demonstrate the superior performance of \emph{LoReKT} against a wide range of recently proposed DLKT models in terms of AUC and Accuracy.

%% file: elsarticle-template-num.bbl
\begin{thebibliography}{62}
\expandafter\ifx\csname natexlab\endcsname\relax\def\natexlab#1{#1}\fi
\providecommand{\url}[1]{\texttt{#1}}
\providecommand{\href}[2]{#2}
\providecommand{\path}[1]{#1}
\providecommand{\DOIprefix}{doi:}
\providecommand{\ArXivprefix}{arXiv:}
\providecommand{\URLprefix}{URL: }
\providecommand{\Pubmedprefix}{pmid:}
\providecommand{\doi}[1]{\href{http://dx.doi.org/#1}{\path{#1}}}
\providecommand{\Pubmed}[1]{\href{pmid:#1}{\path{#1}}}
\providecommand{\bibinfo}[2]{#2}
\ifx\xfnm\relax \def\xfnm[#1]{\unskip,\space#1}\fi
\bibitem[{Zhang and Yao(2018)}]{zhang2018three}
\bibinfo{author}{K.~Zhang}, \bibinfo{author}{Y.~Yao},
\newblock \bibinfo{title}{A three learning states bayesian knowledge tracing
  model},
\newblock \bibinfo{journal}{Knowledge-Based Systems} \bibinfo{volume}{148}
  (\bibinfo{year}{2018}) \bibinfo{pages}{189--201}.
\bibitem[{Su et~al.(2021)Su, Cheng, Luo, Wu, Zhang, Liu, and Wang}]{su2021time}
\bibinfo{author}{Y.~Su}, \bibinfo{author}{Z.~Cheng}, \bibinfo{author}{P.~Luo},
  \bibinfo{author}{J.~Wu}, \bibinfo{author}{L.~Zhang},
  \bibinfo{author}{Q.~Liu}, \bibinfo{author}{S.~Wang},
\newblock \bibinfo{title}{Time-and-concept enhanced deep multidimensional item
  response theory for interpretable knowledge tracing},
\newblock \bibinfo{journal}{Knowledge-Based Systems} \bibinfo{volume}{218}
  (\bibinfo{year}{2021}) \bibinfo{pages}{106819}.
\bibitem[{Song et~al.(2022)Song, Li, Lei, Zhao, Chen, and Mian}]{song2022bi}
\bibinfo{author}{X.~Song}, \bibinfo{author}{J.~Li}, \bibinfo{author}{Q.~Lei},
  \bibinfo{author}{W.~Zhao}, \bibinfo{author}{Y.~Chen},
  \bibinfo{author}{A.~Mian},
\newblock \bibinfo{title}{Bi-clkt: Bi-graph contrastive learning based
  knowledge tracing},
\newblock \bibinfo{journal}{Knowledge-Based Systems} \bibinfo{volume}{241}
  (\bibinfo{year}{2022}) \bibinfo{pages}{108274}.
\bibitem[{Ke et~al.(2024)Ke, Wang, Tan, Du, Jin, Huang, and Yin}]{ke2024hitskt}
\bibinfo{author}{F.~Ke}, \bibinfo{author}{W.~Wang}, \bibinfo{author}{W.~Tan},
  \bibinfo{author}{L.~Du}, \bibinfo{author}{Y.~Jin},
  \bibinfo{author}{Y.~Huang}, \bibinfo{author}{H.~Yin},
\newblock \bibinfo{title}{Hitskt: A hierarchical transformer model for
  session-aware knowledge tracing},
\newblock \bibinfo{journal}{Knowledge-Based Systems} \bibinfo{volume}{284}
  (\bibinfo{year}{2024}) \bibinfo{pages}{111300}.
\bibitem[{Liu et~al.(2019)Liu, Huang, Yin, Chen, Xiong, Su, and
  Hu}]{liu2019ekt}
\bibinfo{author}{Q.~Liu}, \bibinfo{author}{Z.~Huang}, \bibinfo{author}{Y.~Yin},
  \bibinfo{author}{E.~Chen}, \bibinfo{author}{H.~Xiong},
  \bibinfo{author}{Y.~Su}, \bibinfo{author}{G.~Hu},
\newblock \bibinfo{title}{{EKT}: Exercise-aware knowledge tracing for student
  performance prediction},
\newblock \bibinfo{journal}{IEEE Transactions on Knowledge and Data
  Engineering} \bibinfo{volume}{33} (\bibinfo{year}{2019})
  \bibinfo{pages}{100--115}.
\bibitem[{Wu et~al.(2020)Wu, Li, Tang, and Liang}]{wu2020exercise}
\bibinfo{author}{Z.~Wu}, \bibinfo{author}{M.~Li}, \bibinfo{author}{Y.~Tang},
  \bibinfo{author}{Q.~Liang},
\newblock \bibinfo{title}{Exercise recommendation based on knowledge concept
  prediction},
\newblock \bibinfo{journal}{Knowledge-Based Systems} \bibinfo{volume}{210}
  (\bibinfo{year}{2020}) \bibinfo{pages}{106481}.
\bibitem[{Zhang et~al.(2017)Zhang, Shi, King, and Yeung}]{zhang2017dynamic}
\bibinfo{author}{J.~Zhang}, \bibinfo{author}{X.~Shi},
  \bibinfo{author}{I.~King}, \bibinfo{author}{D.~Y. Yeung},
\newblock \bibinfo{title}{Dynamic key-value memory networks for knowledge
  tracing},
\newblock in: \bibinfo{booktitle}{Proceedings of the 26th International
  Conference on World Wide Web}, \bibinfo{year}{2017}, p. \bibinfo{pages}{765}.
\bibitem[{Nagatani et~al.(2019)Nagatani, Zhang, Sato, Chen, Chen, and
  Ohkuma}]{nagatani2019augmenting}
\bibinfo{author}{K.~Nagatani}, \bibinfo{author}{Q.~Zhang},
  \bibinfo{author}{M.~Sato}, \bibinfo{author}{Y.-Y. Chen},
  \bibinfo{author}{F.~Chen}, \bibinfo{author}{T.~Ohkuma},
\newblock \bibinfo{title}{Augmenting knowledge tracing by considering
  forgetting behavior},
\newblock in: \bibinfo{booktitle}{The World Wide Web Conference},
  \bibinfo{year}{2019}, pp. \bibinfo{pages}{3101--3107}.
\bibitem[{Sonkar et~al.(2020)Sonkar, Waters, Lan, Grimaldi, and
  Baraniuk}]{sonkar2020qdkt}
\bibinfo{author}{S.~Sonkar}, \bibinfo{author}{A.~E. Waters},
  \bibinfo{author}{A.~S. Lan}, \bibinfo{author}{P.~J. Grimaldi},
  \bibinfo{author}{R.~G. Baraniuk},
\newblock \bibinfo{title}{q{DKT}: Question-centric deep knowledge tracing},
\newblock in: \bibinfo{booktitle}{Proceedings of The 13th International
  Conference on Educational Data Mining}, \bibinfo{year}{2020}, pp.
  \bibinfo{pages}{677--681}.
\bibitem[{Guo et~al.(2021)Guo, Huang, Gao, Shang, Shu, and
  Sun}]{guo2021enhancing}
\bibinfo{author}{X.~Guo}, \bibinfo{author}{Z.~Huang}, \bibinfo{author}{J.~Gao},
  \bibinfo{author}{M.~Shang}, \bibinfo{author}{M.~Shu},
  \bibinfo{author}{J.~Sun},
\newblock \bibinfo{title}{Enhancing knowledge tracing via adversarial
  training},
\newblock in: \bibinfo{booktitle}{Proceedings of the 29th ACM International
  Conference on Multimedia}, \bibinfo{year}{2021}, pp.
  \bibinfo{pages}{367--375}.
\bibitem[{Zoph et~al.(2016)Zoph, Yuret, May, and Knight}]{zoph2016transfer}
\bibinfo{author}{B.~Zoph}, \bibinfo{author}{D.~Yuret},
  \bibinfo{author}{J.~May}, \bibinfo{author}{K.~Knight},
\newblock \bibinfo{title}{Transfer learning for low-resource neural machine
  translation},
\newblock \bibinfo{journal}{arXiv preprint arXiv:1604.02201}
  (\bibinfo{year}{2016}).
\bibitem[{Tu et~al.(2019)Tu, Chen, Yeh, and Lee}]{tu2019end}
\bibinfo{author}{T.~Tu}, \bibinfo{author}{Y.-J. Chen}, \bibinfo{author}{C.-c.
  Yeh}, \bibinfo{author}{H.-Y. Lee},
\newblock \bibinfo{title}{End-to-end text-to-speech for low-resource languages
  by cross-lingual transfer learning},
\newblock \bibinfo{journal}{arXiv preprint arXiv:1904.06508}
  (\bibinfo{year}{2019}).
\bibitem[{Brown et~al.(2020)Brown, Mann, Ryder, Subbiah, Kaplan, Dhariwal,
  Neelakantan, Shyam, Sastry, Askell et~al.}]{brown2020language}
\bibinfo{author}{T.~Brown}, \bibinfo{author}{B.~Mann},
  \bibinfo{author}{N.~Ryder}, \bibinfo{author}{M.~Subbiah},
  \bibinfo{author}{J.~D. Kaplan}, \bibinfo{author}{P.~Dhariwal},
  \bibinfo{author}{A.~Neelakantan}, \bibinfo{author}{P.~Shyam},
  \bibinfo{author}{G.~Sastry}, \bibinfo{author}{A.~Askell}, et~al.,
\newblock \bibinfo{title}{Language models are few-shot learners},
\newblock \bibinfo{journal}{Advances in neural information processing systems}
  (\bibinfo{year}{2020}) \bibinfo{pages}{1877--1901}.
\bibitem[{Gira et~al.(2022)Gira, Zhang, and Lee}]{gira2022debiasing}
\bibinfo{author}{M.~Gira}, \bibinfo{author}{R.~Zhang},
  \bibinfo{author}{K.~Lee},
\newblock \bibinfo{title}{Debiasing pre-trained language models via efficient
  fine-tuning},
\newblock in: \bibinfo{booktitle}{Proceedings of the Second Workshop on
  Language Technology for Equality, Diversity and Inclusion},
  \bibinfo{year}{2022}, pp. \bibinfo{pages}{59--69}.
\bibitem[{Liu et~al.(2022)Liu, Liu, Chen, Huang, Tang, and Luo}]{liu2022pykt}
\bibinfo{author}{Z.~Liu}, \bibinfo{author}{Q.~Liu}, \bibinfo{author}{J.~Chen},
  \bibinfo{author}{S.~Huang}, \bibinfo{author}{J.~Tang},
  \bibinfo{author}{W.~Luo},
\newblock \bibinfo{title}{\textsc{pyKT}: A python library to benchmark deep
  learning based knowledge tracing models},
\newblock in: \bibinfo{booktitle}{Thirty-sixth Conference on Neural Information
  Processing Systems}, \bibinfo{year}{2022}.
\bibitem[{Piech et~al.(2015)Piech, Bassen, Huang, Ganguli, Sahami, Guibas, and
  Sohl-Dickstein}]{piech2015deep}
\bibinfo{author}{C.~Piech}, \bibinfo{author}{J.~Bassen},
  \bibinfo{author}{J.~Huang}, \bibinfo{author}{S.~Ganguli},
  \bibinfo{author}{M.~Sahami}, \bibinfo{author}{L.~J. Guibas},
  \bibinfo{author}{J.~Sohl-Dickstein},
\newblock \bibinfo{title}{Deep knowledge tracing},
\newblock \bibinfo{journal}{Advances in neural information processing systems}
  \bibinfo{volume}{28} (\bibinfo{year}{2015}).
\bibitem[{Yeung and Yeung(2018)}]{yeung2018addressing}
\bibinfo{author}{C.-K. Yeung}, \bibinfo{author}{D.-Y. Yeung},
\newblock \bibinfo{title}{Addressing two problems in deep knowledge tracing via
  prediction-consistent regularization},
\newblock in: \bibinfo{booktitle}{Proceedings of the Fifth Annual ACM
  Conference on Learning at Scale}, \bibinfo{year}{2018}, pp.
  \bibinfo{pages}{1--10}.
\bibitem[{Lee and Yeung(2019)}]{lee2019knowledge}
\bibinfo{author}{J.~Lee}, \bibinfo{author}{D.-Y. Yeung},
\newblock \bibinfo{title}{Knowledge query network for knowledge tracing: How
  knowledge interacts with skills},
\newblock in: \bibinfo{booktitle}{Proceedings of the 9th International
  Conference on Learning Analytics \& Knowledge}, \bibinfo{year}{2019}, pp.
  \bibinfo{pages}{491--500}.
\bibitem[{Pandey and Karypis(2019)}]{pandey2019self}
\bibinfo{author}{S.~Pandey}, \bibinfo{author}{G.~Karypis},
\newblock \bibinfo{title}{A self-attentive model for knowledge tracing},
\newblock in: \bibinfo{booktitle}{12th International Conference on Educational
  Data Mining}, \bibinfo{organization}{International Educational Data Mining
  Society}, \bibinfo{year}{2019}, pp. \bibinfo{pages}{384--389}.
\bibitem[{Ghosh et~al.(2020)Ghosh, Heffernan, and Lan}]{ghosh2020context}
\bibinfo{author}{A.~Ghosh}, \bibinfo{author}{N.~Heffernan},
  \bibinfo{author}{A.~S. Lan},
\newblock \bibinfo{title}{Context-aware attentive knowledge tracing},
\newblock in: \bibinfo{booktitle}{ACM SIGKDD Conference on Knowledge Discovery
  and Data Mining}, \bibinfo{year}{2020}.
\bibitem[{Tiana et~al.(2021)Tiana, Zhengc, Flanaganb, Mic, and
  Ogatab}]{tiana2021bekt}
\bibinfo{author}{Z.~Tiana}, \bibinfo{author}{G.~Zhengc},
  \bibinfo{author}{B.~Flanaganb}, \bibinfo{author}{J.~Mic},
  \bibinfo{author}{H.~Ogatab},
\newblock \bibinfo{title}{Bekt: Deep knowledge tracing with bidirectional
  encoder representations from transformers},
\newblock in: \bibinfo{booktitle}{Proceedings of the 29th International
  Conference on Computers in Education}, \bibinfo{year}{2021}.
\bibitem[{Song et~al.(2022)Song, Li, Cai, Yang, Yang, and Liu}]{song2022survey}
\bibinfo{author}{X.~Song}, \bibinfo{author}{J.~Li}, \bibinfo{author}{T.~Cai},
  \bibinfo{author}{S.~Yang}, \bibinfo{author}{T.~Yang},
  \bibinfo{author}{C.~Liu},
\newblock \bibinfo{title}{A survey on deep learning based knowledge tracing},
\newblock \bibinfo{journal}{Knowledge-Based Systems} \bibinfo{volume}{258}
  (\bibinfo{year}{2022}) \bibinfo{pages}{110036}.
\bibitem[{Ma et~al.(2022)Ma, Han, Qiao, Cui, Yin, and Yu}]{ma2022spakt}
\bibinfo{author}{Y.~Ma}, \bibinfo{author}{P.~Han}, \bibinfo{author}{H.~Qiao},
  \bibinfo{author}{C.~Cui}, \bibinfo{author}{Y.~Yin}, \bibinfo{author}{D.~Yu},
\newblock \bibinfo{title}{Spakt: A self-supervised pre-training method for
  knowledge tracing},
\newblock \bibinfo{journal}{IEEE Access} \bibinfo{volume}{10}
  (\bibinfo{year}{2022}) \bibinfo{pages}{72145--72154}.
\bibitem[{Liu et~al.(2023)Liu, Liu, Chen, Huang, and Luo}]{liu2023simplekt}
\bibinfo{author}{Z.~Liu}, \bibinfo{author}{Q.~Liu}, \bibinfo{author}{J.~Chen},
  \bibinfo{author}{S.~Huang}, \bibinfo{author}{W.~Luo},
\newblock \bibinfo{title}{simple{KT}: A simple but tough-to-beat baseline for
  knowledge tracing},
\newblock in: \bibinfo{booktitle}{International Conference on Learning
  Representations}, \bibinfo{year}{2023}.
\bibitem[{Zhang et~al.(2024)Zhang, Liu, Shang, Li, and
  Jiang}]{zhang2024question}
\bibinfo{author}{H.~Zhang}, \bibinfo{author}{Z.~Liu},
  \bibinfo{author}{C.~Shang}, \bibinfo{author}{D.~Li},
  \bibinfo{author}{Y.~Jiang},
\newblock \bibinfo{title}{A question-centric multi-experts contrastive learning
  framework for improving the accuracy and interpretability of deep sequential
  knowledge tracing models},
\newblock \bibinfo{journal}{arXiv preprint arXiv:2403.07322}
  (\bibinfo{year}{2024}).
\bibitem[{Murray et~al.(2013)Murray, Ritter, Nixon, Schwiebert, Hausmann,
  Towle, Fancsali, and Vuong}]{learning_curve}
\bibinfo{author}{R.~C. Murray}, \bibinfo{author}{S.~Ritter},
  \bibinfo{author}{T.~Nixon}, \bibinfo{author}{R.~Schwiebert},
  \bibinfo{author}{R.~G. Hausmann}, \bibinfo{author}{B.~Towle},
  \bibinfo{author}{S.~E. Fancsali}, \bibinfo{author}{A.~Vuong},
\newblock \bibinfo{title}{Revealing the learning in learning curves},
\newblock in: \bibinfo{booktitle}{International Conference on Artificial
  Intelligence in Education}, \bibinfo{organization}{Springer},
  \bibinfo{year}{2013}, pp. \bibinfo{pages}{473--482}.
\bibitem[{Nakagawa et~al.(2019)Nakagawa, Iwasawa, and
  Matsuo}]{nakagawa2019graph}
\bibinfo{author}{H.~Nakagawa}, \bibinfo{author}{Y.~Iwasawa},
  \bibinfo{author}{Y.~Matsuo},
\newblock \bibinfo{title}{Graph-based knowledge tracing: modeling student
  proficiency using graph neural network},
\newblock in: \bibinfo{booktitle}{2019 IEEE/WIC/ACM International Conference on
  Web Intelligence}, \bibinfo{organization}{IEEE}, \bibinfo{year}{2019}, pp.
  \bibinfo{pages}{156--163}.
\bibitem[{Yosinski et~al.(2014)Yosinski, Clune, Bengio, and
  Lipson}]{yosinski2014transferable}
\bibinfo{author}{J.~Yosinski}, \bibinfo{author}{J.~Clune},
  \bibinfo{author}{Y.~Bengio}, \bibinfo{author}{H.~Lipson},
\newblock \bibinfo{title}{How transferable are features in deep neural
  networks?},
\newblock \bibinfo{journal}{Advances in neural information processing systems}
  \bibinfo{volume}{27} (\bibinfo{year}{2014}).
\bibitem[{Shang et~al.(2019)Shang, Ma, Xiao, and Sun}]{shang2019pre}
\bibinfo{author}{J.~Shang}, \bibinfo{author}{T.~Ma}, \bibinfo{author}{C.~Xiao},
  \bibinfo{author}{J.~Sun},
\newblock \bibinfo{title}{Pre-training of graph augmented transformers for
  medication recommendation},
\newblock \bibinfo{journal}{arXiv preprint arXiv:1906.00346}
  (\bibinfo{year}{2019}).
\bibitem[{Bansal et~al.(2019)Bansal, Kamper, Livescu, Lopez, and
  Goldwater}]{bansal2019pre}
\bibinfo{author}{S.~Bansal}, \bibinfo{author}{H.~Kamper},
  \bibinfo{author}{K.~Livescu}, \bibinfo{author}{A.~Lopez},
  \bibinfo{author}{S.~Goldwater},
\newblock \bibinfo{title}{Pre-training on high-resource speech recognition
  improves low-resource speech-to-text translation},
\newblock in: \bibinfo{booktitle}{2019 Annual Conference of the North American
  Chapter of the Association for Computational Linguistics},
  \bibinfo{organization}{Association for Computational Linguistics},
  \bibinfo{year}{2019}, pp. \bibinfo{pages}{58--68}.
\bibitem[{Zhang et~al.(2021)Zhang, Xia, Yu, Liu, and Zhao}]{zhang2021pdaln}
\bibinfo{author}{T.~Zhang}, \bibinfo{author}{C.~Xia}, \bibinfo{author}{P.~S.
  Yu}, \bibinfo{author}{Z.~Liu}, \bibinfo{author}{S.~Zhao},
\newblock \bibinfo{title}{Pdaln: Progressive domain adaptation over a
  pre-trained model for low-resource cross-domain named entity recognition},
\newblock in: \bibinfo{booktitle}{EMNLP}, \bibinfo{year}{2021}.
\bibitem[{Liu et~al.(2022)Liu, Xu, Xu, Qian, Li, Ji, Chan, and
  Jin}]{liu2022improved}
\bibinfo{author}{Z.~Liu}, \bibinfo{author}{Y.~Xu}, \bibinfo{author}{Y.~Xu},
  \bibinfo{author}{Q.~Qian}, \bibinfo{author}{H.~Li}, \bibinfo{author}{X.~Ji},
  \bibinfo{author}{A.~Chan}, \bibinfo{author}{R.~Jin},
\newblock \bibinfo{title}{Improved fine-tuning by better leveraging
  pre-training data},
\newblock \bibinfo{journal}{Advances in Neural Information Processing Systems}
  \bibinfo{volume}{35} (\bibinfo{year}{2022}) \bibinfo{pages}{32568--32581}.
\bibitem[{Chi et~al.(2023)Chi, Huang, Liu, Bai, Gao, and Mao}]{10103146}
\bibinfo{author}{Z.~Chi}, \bibinfo{author}{H.~Huang}, \bibinfo{author}{L.~Liu},
  \bibinfo{author}{Y.~Bai}, \bibinfo{author}{X.~Gao}, \bibinfo{author}{X.-L.
  Mao},
\newblock \bibinfo{title}{Can pretrained english language models benefit
  non-english nlp systems in low-resource scenarios?},
\newblock \bibinfo{journal}{IEEE/ACM Transactions on Audio, Speech, and
  Language Processing}  (\bibinfo{year}{2023}) \bibinfo{pages}{1--14}.
  \DOIprefix\doi{10.1109/TASLP.2023.3267618}.
\bibitem[{Zhang et~al.(2023)Zhang, Li, Li, Shang, Shi, and
  Jiang}]{zhang-etal-2023-assisting}
\bibinfo{author}{H.~Zhang}, \bibinfo{author}{D.~Li}, \bibinfo{author}{Y.~Li},
  \bibinfo{author}{C.~Shang}, \bibinfo{author}{C.~Shi},
  \bibinfo{author}{Y.~Jiang},
\newblock \bibinfo{title}{Assisting language learners: Automated trans-lingual
  definition generation via contrastive prompt learning},
\newblock in: \bibinfo{booktitle}{Proceedings of the 18th Workshop on
  Innovative Use of NLP for Building Educational Applications (BEA 2023)},
  \bibinfo{year}{2023}, pp. \bibinfo{pages}{260--274}.
\bibitem[{Yeung and Yeung(2018)}]{dkt+}
\bibinfo{author}{C.-K. Yeung}, \bibinfo{author}{D.-Y. Yeung},
\newblock \bibinfo{title}{Addressing two problems in deep knowledge tracing via
  prediction-consistent regularization},
\newblock in: \bibinfo{booktitle}{Proceedings of the Fifth Annual ACM
  Conference on Learning at Scale}, \bibinfo{year}{2018}, pp.
  \bibinfo{pages}{1--10}.
\bibitem[{Nagatani et~al.(2019)Nagatani, Zhang, Sato, Chen, Chen, and
  Ohkuma}]{dkt-forget}
\bibinfo{author}{K.~Nagatani}, \bibinfo{author}{Q.~Zhang},
  \bibinfo{author}{M.~Sato}, \bibinfo{author}{Y.-Y. Chen},
  \bibinfo{author}{F.~Chen}, \bibinfo{author}{T.~Ohkuma},
\newblock \bibinfo{title}{Augmenting knowledge tracing by considering
  forgetting behavior},
\newblock in: \bibinfo{booktitle}{The world wide web conference},
  \bibinfo{year}{2019}, pp. \bibinfo{pages}{3101--3107}.
\bibitem[{Choi et~al.(2020)Choi, Lee, Cho, Baek, Kim, Cha, Shin, Bae, and
  Heo}]{choi2020towards}
\bibinfo{author}{Y.~Choi}, \bibinfo{author}{Y.~Lee}, \bibinfo{author}{J.~Cho},
  \bibinfo{author}{J.~Baek}, \bibinfo{author}{B.~Kim},
  \bibinfo{author}{Y.~Cha}, \bibinfo{author}{D.~Shin},
  \bibinfo{author}{C.~Bae}, \bibinfo{author}{J.~Heo},
\newblock \bibinfo{title}{Towards an appropriate query, key, and value
  computation for knowledge tracing},
\newblock in: \bibinfo{booktitle}{Proceedings of the Seventh ACM Conference on
  Learning@Scale}, \bibinfo{year}{2020}, pp. \bibinfo{pages}{341--344}.
\bibitem[{Devlin et~al.(2019)Devlin, Chang, Lee, and Toutanova}]{bert}
\bibinfo{author}{J.~Devlin}, \bibinfo{author}{M.-W. Chang},
  \bibinfo{author}{K.~Lee}, \bibinfo{author}{K.~Toutanova},
\newblock \bibinfo{title}{{BERT}: Pre-training of deep bidirectional
  transformers for language understanding},
\newblock in: \bibinfo{booktitle}{Proceedings of the 2019 Conference of the
  North {A}merican Chapter of the Association for Computational Linguistics},
  \bibinfo{year}{2019}, pp. \bibinfo{pages}{4171--4186}.
\bibitem[{Radford et~al.(2018)Radford, Narasimhan, Salimans, Sutskever
  et~al.}]{radford2018improving}
\bibinfo{author}{A.~Radford}, \bibinfo{author}{K.~Narasimhan},
  \bibinfo{author}{T.~Salimans}, \bibinfo{author}{I.~Sutskever}, et~al.,
\newblock \bibinfo{title}{Improving language understanding by generative
  pre-training}  (\bibinfo{year}{2018}).
\bibitem[{Raffel et~al.(2020)Raffel, Shazeer, Roberts, Lee, Narang, Matena,
  Zhou, Li, and Liu}]{raffel2020exploring}
\bibinfo{author}{C.~Raffel}, \bibinfo{author}{N.~Shazeer},
  \bibinfo{author}{A.~Roberts}, \bibinfo{author}{K.~Lee},
  \bibinfo{author}{S.~Narang}, \bibinfo{author}{M.~Matena},
  \bibinfo{author}{Y.~Zhou}, \bibinfo{author}{W.~Li}, \bibinfo{author}{P.~J.
  Liu},
\newblock \bibinfo{title}{Exploring the limits of transfer learning with a
  unified text-to-text transformer},
\newblock \bibinfo{journal}{The Journal of Machine Learning Research}
  \bibinfo{volume}{21} (\bibinfo{year}{2020}) \bibinfo{pages}{5485--5551}.
\bibitem[{Lester et~al.(2021)Lester, Al-Rfou, and
  Constant}]{lester-etal-2021-power}
\bibinfo{author}{B.~Lester}, \bibinfo{author}{R.~Al-Rfou},
  \bibinfo{author}{N.~Constant},
\newblock \bibinfo{title}{The power of scale for parameter-efficient prompt
  tuning},
\newblock in: \bibinfo{editor}{M.-F. Moens}, \bibinfo{editor}{X.~Huang},
  \bibinfo{editor}{L.~Specia}, \bibinfo{editor}{S.~W.-t. Yih} (Eds.),
  \bibinfo{booktitle}{Proceedings of the 2021 Conference on Empirical Methods
  in Natural Language Processing}, \bibinfo{publisher}{Association for
  Computational Linguistics}, \bibinfo{address}{Online and Punta Cana,
  Dominican Republic}, \bibinfo{year}{2021}, pp. \bibinfo{pages}{3045--3059}.
  \URLprefix \url{https://aclanthology.org/2021.emnlp-main.243}.
  \DOIprefix\doi{10.18653/v1/2021.emnlp-main.243}.
\bibitem[{Sanh et~al.(2022)Sanh, Webson, Raffel, Bach, Sutawika, Alyafeai,
  Chaffin, Stiegler, Raja, Dey, Bari, Xu, Thakker, Sharma, Szczechla, Kim,
  Chhablani, Nayak, Datta, Chang, Jiang, Wang, Manica, Shen, Yong, Pandey,
  Bawden, Wang, Neeraj, Rozen, Sharma, Santilli, Fevry, Fries, Teehan, Scao,
  Biderman, Gao, Wolf, and Rush}]{sanh2021multitask}
\bibinfo{author}{V.~Sanh}, \bibinfo{author}{A.~Webson},
  \bibinfo{author}{C.~Raffel}, \bibinfo{author}{S.~Bach},
  \bibinfo{author}{L.~Sutawika}, \bibinfo{author}{Z.~Alyafeai},
  \bibinfo{author}{A.~Chaffin}, \bibinfo{author}{A.~Stiegler},
  \bibinfo{author}{A.~Raja}, \bibinfo{author}{M.~Dey}, \bibinfo{author}{M.~S.
  Bari}, \bibinfo{author}{C.~Xu}, \bibinfo{author}{U.~Thakker},
  \bibinfo{author}{S.~S. Sharma}, \bibinfo{author}{E.~Szczechla},
  \bibinfo{author}{T.~Kim}, \bibinfo{author}{G.~Chhablani},
  \bibinfo{author}{N.~Nayak}, \bibinfo{author}{D.~Datta},
  \bibinfo{author}{J.~Chang}, \bibinfo{author}{M.~T.-J. Jiang},
  \bibinfo{author}{H.~Wang}, \bibinfo{author}{M.~Manica},
  \bibinfo{author}{S.~Shen}, \bibinfo{author}{Z.~X. Yong},
  \bibinfo{author}{H.~Pandey}, \bibinfo{author}{R.~Bawden},
  \bibinfo{author}{T.~Wang}, \bibinfo{author}{T.~Neeraj},
  \bibinfo{author}{J.~Rozen}, \bibinfo{author}{A.~Sharma},
  \bibinfo{author}{A.~Santilli}, \bibinfo{author}{T.~Fevry},
  \bibinfo{author}{J.~A. Fries}, \bibinfo{author}{R.~Teehan},
  \bibinfo{author}{T.~L. Scao}, \bibinfo{author}{S.~Biderman},
  \bibinfo{author}{L.~Gao}, \bibinfo{author}{T.~Wolf}, \bibinfo{author}{A.~M.
  Rush},
\newblock \bibinfo{title}{Multitask prompted training enables zero-shot task
  generalization},
\newblock in: \bibinfo{booktitle}{International Conference on Learning
  Representations}, \bibinfo{year}{2022}.
\bibitem[{Chung et~al.(2022)Chung, Hou, Longpre, Zoph, Tay, Fedus, Li, Wang,
  Dehghani, Brahma et~al.}]{chung2022scaling}
\bibinfo{author}{H.~W. Chung}, \bibinfo{author}{L.~Hou},
  \bibinfo{author}{S.~Longpre}, \bibinfo{author}{B.~Zoph},
  \bibinfo{author}{Y.~Tay}, \bibinfo{author}{W.~Fedus},
  \bibinfo{author}{Y.~Li}, \bibinfo{author}{X.~Wang},
  \bibinfo{author}{M.~Dehghani}, \bibinfo{author}{S.~Brahma}, et~al.,
\newblock \bibinfo{title}{Scaling instruction-finetuned language models},
\newblock \bibinfo{journal}{CoRR}  (\bibinfo{year}{2022}).
\bibitem[{Michel et~al.(2019)Michel, Levy, and Neubig}]{michel2019sixteen}
\bibinfo{author}{P.~Michel}, \bibinfo{author}{O.~Levy},
  \bibinfo{author}{G.~Neubig},
\newblock \bibinfo{title}{Are sixteen heads really better than one?},
\newblock \bibinfo{journal}{Advances in neural information processing systems}
  \bibinfo{volume}{32} (\bibinfo{year}{2019}).
\bibitem[{Ke et~al.(2022)Ke, Shao, Lin, Konishi, Kim, and
  Liu}]{ke2022continual}
\bibinfo{author}{Z.~Ke}, \bibinfo{author}{Y.~Shao}, \bibinfo{author}{H.~Lin},
  \bibinfo{author}{T.~Konishi}, \bibinfo{author}{G.~Kim},
  \bibinfo{author}{B.~Liu},
\newblock \bibinfo{title}{Continual pre-training of language models},
\newblock in: \bibinfo{booktitle}{The Eleventh International Conference on
  Learning Representations}, \bibinfo{year}{2022}.
\bibitem[{Kong et~al.(2022)Kong, Chen, Zhang, Yang, and
  Yang}]{kong2022multitasking}
\bibinfo{author}{C.~Kong}, \bibinfo{author}{Y.~Chen},
  \bibinfo{author}{H.~Zhang}, \bibinfo{author}{L.~Yang},
  \bibinfo{author}{E.~Yang},
\newblock \bibinfo{title}{Multitasking framework for unsupervised simple
  definition generation},
\newblock \bibinfo{journal}{arXiv preprint arXiv:2203.12926}
  (\bibinfo{year}{2022}).
\bibitem[{Li et~al.(2023)Li, Zhang, Li, and Yang}]{li2023multi}
\bibinfo{author}{D.~Li}, \bibinfo{author}{H.~Zhang}, \bibinfo{author}{Y.~Li},
  \bibinfo{author}{S.~Yang},
\newblock \bibinfo{title}{Multi-level contrastive learning for script-based
  character understanding},
\newblock \bibinfo{journal}{arXiv preprint arXiv:2310.13231}
  (\bibinfo{year}{2023}).
\bibitem[{Stamper et~al.(2010)Stamper, Niculescu-Mizil, Ritter, Gordon, and
  Koedinger}]{algebra}
\bibinfo{author}{J.~Stamper}, \bibinfo{author}{A.~Niculescu-Mizil},
  \bibinfo{author}{S.~Ritter}, \bibinfo{author}{G.~Gordon},
  \bibinfo{author}{K.~Koedinger},
\newblock \bibinfo{title}{Challenge data set from kdd cup 2010 educational data
  mining challenge}  (\bibinfo{year}{2010}).
\bibitem[{Liu et~al.(2023)Liu, Liu, Guo, Chen, Huang, Zhao, Tang, Luo, and
  Weng}]{liu2023xes3g5m}
\bibinfo{author}{Z.~Liu}, \bibinfo{author}{Q.~Liu}, \bibinfo{author}{T.~Guo},
  \bibinfo{author}{J.~Chen}, \bibinfo{author}{S.~Huang},
  \bibinfo{author}{X.~Zhao}, \bibinfo{author}{J.~Tang},
  \bibinfo{author}{W.~Luo}, \bibinfo{author}{J.~Weng},
\newblock \bibinfo{title}{Xes3g5m: A knowledge tracing benchmark dataset with
  auxiliary information},
\newblock in: \bibinfo{booktitle}{Thirty-seventh Conference on Neural
  Information Processing Systems Datasets and Benchmarks Track},
  \bibinfo{year}{2023}.
\bibitem[{Choi et~al.(2020)Choi, Lee, Shin, Cho, Park, Lee, Baek, Bae, Kim, and
  Heo}]{ednet}
\bibinfo{author}{Y.~Choi}, \bibinfo{author}{Y.~Lee}, \bibinfo{author}{D.~Shin},
  \bibinfo{author}{J.~Cho}, \bibinfo{author}{S.~Park},
  \bibinfo{author}{S.~Lee}, \bibinfo{author}{J.~Baek},
  \bibinfo{author}{C.~Bae}, \bibinfo{author}{B.~Kim}, \bibinfo{author}{J.~Heo},
\newblock \bibinfo{title}{Ednet: A large-scale hierarchical dataset in
  education},
\newblock in: \bibinfo{booktitle}{Artificial Intelligence in Education: 21st
  International Conference, AIED 2020, Ifrane, Morocco, July 6--10, 2020,
  Proceedings, Part II 21}, \bibinfo{organization}{Springer},
  \bibinfo{year}{2020}, pp. \bibinfo{pages}{69--73}.
\bibitem[{Feng et~al.(2009)Feng, Heffernan, and Koedinger}]{as2009}
\bibinfo{author}{M.~Feng}, \bibinfo{author}{N.~Heffernan},
  \bibinfo{author}{K.~Koedinger},
\newblock \bibinfo{title}{Addressing the assessment challenge with an online
  system that tutors as it assesses},
\newblock \bibinfo{journal}{User modeling and user-adapted interaction}
  \bibinfo{volume}{19} (\bibinfo{year}{2009}) \bibinfo{pages}{243--266}.
\bibitem[{Wang et~al.(2020)Wang, Lamb, Saveliev, Cameron, Zaykov,
  Hern{\'a}ndez-Lobato, Turner, Baraniuk, Barton, Jones et~al.}]{nips}
\bibinfo{author}{Z.~Wang}, \bibinfo{author}{A.~Lamb},
  \bibinfo{author}{E.~Saveliev}, \bibinfo{author}{P.~Cameron},
  \bibinfo{author}{Y.~Zaykov}, \bibinfo{author}{J.~M. Hern{\'a}ndez-Lobato},
  \bibinfo{author}{R.~E. Turner}, \bibinfo{author}{R.~G. Baraniuk},
  \bibinfo{author}{C.~Barton}, \bibinfo{author}{S.~P. Jones}, et~al.,
\newblock \bibinfo{title}{Instructions and guide for diagnostic questions: The
  neurips 2020 education challenge},
\newblock \bibinfo{journal}{arXiv preprint arXiv:2007.12061}
  (\bibinfo{year}{2020}).
\bibitem[{Kingma and Ba(2015)}]{kingma2015adam}
\bibinfo{author}{D.~P. Kingma}, \bibinfo{author}{J.~Ba},
\newblock \bibinfo{title}{Adam: A method for stochastic optimization},
\newblock in: \bibinfo{booktitle}{International Conference on Learning
  Representations}, \bibinfo{year}{2015}.
\bibitem[{Shen et~al.(2021)Shen, Liu, Chen, Huang, Huang, Yin, Su, and
  Wang}]{shen2021learning}
\bibinfo{author}{S.~Shen}, \bibinfo{author}{Q.~Liu}, \bibinfo{author}{E.~Chen},
  \bibinfo{author}{Z.~Huang}, \bibinfo{author}{W.~Huang},
  \bibinfo{author}{Y.~Yin}, \bibinfo{author}{Y.~Su}, \bibinfo{author}{S.~Wang},
\newblock \bibinfo{title}{Learning process-consistent knowledge tracing},
\newblock in: \bibinfo{booktitle}{Proceedings of the 27th ACM SIGKDD Conference
  on Knowledge Discovery \& Data Mining}, \bibinfo{year}{2021}, pp.
  \bibinfo{pages}{1452--1460}.
\bibitem[{Liu et~al.(2023)Liu, Liu, Chen, Huang, Gao, Luo, and
  Weng}]{liu2023enhancing}
\bibinfo{author}{Z.~Liu}, \bibinfo{author}{Q.~Liu}, \bibinfo{author}{J.~Chen},
  \bibinfo{author}{S.~Huang}, \bibinfo{author}{B.~Gao},
  \bibinfo{author}{W.~Luo}, \bibinfo{author}{J.~Weng},
\newblock \bibinfo{title}{Enhancing deep knowledge tracing with auxiliary
  tasks},
\newblock in: \bibinfo{booktitle}{Proceedings of the 2023 World Wide Web
  Conference}, \bibinfo{year}{2023}.
\bibitem[{Abdelrahman and Wang(2019)}]{abdelrahman2019knowledge}
\bibinfo{author}{G.~Abdelrahman}, \bibinfo{author}{Q.~Wang},
\newblock \bibinfo{title}{Knowledge tracing with sequential key-value memory
  networks},
\newblock in: \bibinfo{booktitle}{Proceedings of the 42nd International ACM
  SIGIR Conference on Research and Development in Information Retrieval},
  \bibinfo{year}{2019}, pp. \bibinfo{pages}{175--184}.
\bibitem[{Yeung(2019)}]{yeung2019deep}
\bibinfo{author}{C.-K. Yeung},
\newblock \bibinfo{title}{Deep-{IRT}: Make deep learning based knowledge
  tracing explainable using item response theory},
\newblock \bibinfo{journal}{arXiv preprint arXiv:1904.11738}
  (\bibinfo{year}{2019}).
\bibitem[{Huang et~al.(2023)Huang, Liu, Zhao, Luo, and Weng}]{huang2023towards}
\bibinfo{author}{S.~Huang}, \bibinfo{author}{Z.~Liu},
  \bibinfo{author}{X.~Zhao}, \bibinfo{author}{W.~Luo},
  \bibinfo{author}{J.~Weng},
\newblock \bibinfo{title}{Towards robust knowledge tracing models via k-sparse
  attention},
\newblock in: \bibinfo{booktitle}{Proceedings of the 46th International ACM
  SIGIR Conference on Research and Development in Information Retrieval},
  \bibinfo{year}{2023}, pp. \bibinfo{pages}{2441--2445}.
\bibitem[{Wang et~al.(2021)Wang, Ma, Zhang, Lv, Wan, Lin, Tang, Liu, and
  Ma}]{wang2021temporal}
\bibinfo{author}{C.~Wang}, \bibinfo{author}{W.~Ma}, \bibinfo{author}{M.~Zhang},
  \bibinfo{author}{C.~Lv}, \bibinfo{author}{F.~Wan}, \bibinfo{author}{H.~Lin},
  \bibinfo{author}{T.~Tang}, \bibinfo{author}{Y.~Liu}, \bibinfo{author}{S.~Ma},
\newblock \bibinfo{title}{Temporal cross-effects in knowledge tracing},
\newblock in: \bibinfo{booktitle}{Proceedings of the 14th ACM International
  Conference on Web Search and Data Mining}, \bibinfo{year}{2021}, pp.
  \bibinfo{pages}{517--525}.
\bibitem[{Shang et~al.(2024)Shang, Zhang, Wen, and
  Yang}]{shang2024understanding}
\bibinfo{author}{C.~Shang}, \bibinfo{author}{H.~Zhang},
  \bibinfo{author}{H.~Wen}, \bibinfo{author}{Y.~Yang},
\newblock \bibinfo{title}{Understanding multimodal deep neural networks: A
  concept selection view},
\newblock \bibinfo{journal}{arXiv preprint arXiv:2404.08964}
  (\bibinfo{year}{2024}).
\bibitem[{Zhang et~al.(2022)Zhang, Li, Yang, and Li}]{zhang2022fine}
\bibinfo{author}{H.~Zhang}, \bibinfo{author}{D.~Li}, \bibinfo{author}{S.~Yang},
  \bibinfo{author}{Y.~Li},
\newblock \bibinfo{title}{Fine-grained contrastive learning for definition
  generation},
\newblock in: \bibinfo{booktitle}{Proceedings of the 2nd Conference of the
  Asia-Pacific Chapter of the Association for Computational Linguistics and the
  12th International Joint Conference on Natural Language Processing},
  \bibinfo{year}{2022}, pp. \bibinfo{pages}{1001--1012}.
\bibitem[{Kong et~al.(2022)Kong, Wang, Chong, Yang, Zhang, Yang, and
  Huang}]{kong2022blcu}
\bibinfo{author}{C.~Kong}, \bibinfo{author}{Y.~Wang},
  \bibinfo{author}{R.~Chong}, \bibinfo{author}{L.~Yang},
  \bibinfo{author}{H.~Zhang}, \bibinfo{author}{E.~Yang},
  \bibinfo{author}{Y.~Huang},
\newblock \bibinfo{title}{Blcu-icall at semeval-2022 task 1: Cross-attention
  multitasking framework for definition modeling},
\newblock \bibinfo{journal}{arXiv preprint arXiv:2204.07701}
  (\bibinfo{year}{2022}).

\end{thebibliography}
